\documentclass[]{aa}  

\usepackage{amsmath}

\usepackage{txfonts} % STRONGLY recommended by A&A
\usepackage{natbib,twoopt} % for the bibliography
\bibpunct{(}{)}{;}{a}{}{,} % to follow the A&A style
\usepackage{txfonts} % STRONGLY recommended by A&A
\usepackage{graphicx}
\usepackage{epstopdf}
\makeatletter
\newcommandtwoopt{\citeads}[3][][]{\href{http://adsabs.harvard.edu/abs/#3}%
{\def\hyper@linkstart##1##2{}%
\let\hyper@linkend\@empty\citealp[#1][#2]{#3}}}
\newcommand{\Msun}{M$_\odot$}

\newcommand{\abratio}[2]{[\mathrm{#1}/\mathrm{#2}]\xspace}

\begin{document}

   \title{The mass-ratio and eccentricity distributions of  barium and S stars, and red giants in open clusters\thanks{Based on observations made with the Mercator Telescope, operated on the island of La Palma by the Flemish Community, at the Spanish Observatorio del Roque de los Muchachos of the Instituto de Astrofisica de Canarias, and on observations made with the HARPS spectrograph installed on the 3.6~m telescope at the European Southern Observatory.}}
   \titlerunning{Orbital properties of barium and S stars, and red giants in open clusters}
  % \subtitle{SUBTiTLE}

\author{Mathieu Van
der Swaelmen\inst{1}
 \and
   Henri M.J. Boffin \inst{2,3}
   \and
      Alain Jorissen
          \inst{1}
  \and
   Sophie Van Eck 
   \inst{1}
  }

   \institute{
   Institut d'Astronomie et d'Astrophysique, Universit\'e Libre de Bruxelles, Campus Plaine C.P. 226, Bd du Triomphe, B-1050 Bruxelles, Belgium\\
\email{ajorisse,svaneck@ulb.ac.be}
   \and
ESO, Alonso de C\'ordova 3107, Casilla 19001, Santiago, Chile\\
\email{hboffin@eso.org}
\and
ESO, Karl Schwarzschild Strasse 2, 85748 Garching, Germany
%              \and   
%    Instituut voor Sterrenkunde, Katholieke Universiteit Leuven, Celestijnenlaan 200D, %3001 Leuven, Belgium
}

   \date{Received May 9, 2016; accepted August 11, 2016}

 \abstract{A complete set of orbital parameters for barium stars, including the longest orbits, has recently been obtained thanks to a radial-velocity monitoring with the HERMES spectrograph installed on the Flemish Mercator telescope. Barium stars are supposed to belong to post-mass-transfer systems.}
{In order to identify diagnostics distinguishing between pre- and post-mass-transfer systems, the properties of barium stars (more precisely their mass-function distribution and their period -- eccentricity ($P - e$) diagram) are compared to those of  binary red giants in open clusters. As a side product, we aim to identify possible post-mass-transfer systems among the cluster giants from the presence of $s$-process overabundances. We investigate the relation between the $s$-process enrichment, the location in the $(P-e)$ diagram, and the cluster metallicity and turn-off mass.}{To invert the mass-function distribution and derive the mass-ratio distribution, we used the method pioneered by Boffin et al. (1992) that relies on a Richardson-Lucy deconvolution algorithm. The derivation of $s$-process abundances in the open-cluster giants was performed through spectral synthesis with MARCS model atmospheres.}{A fraction of 22\% of post-mass-transfer systems is found among the cluster binary giants (with companion masses between 0.58 and
0.87~M$_\odot$, typical for white dwarfs), and these systems occupy a wider area than barium stars in the
$(P - e)$ diagram. Barium stars have on average lower eccentricities at a given orbital period. When the sample of  binary giant stars in clusters is restricted to the subsample of systems occupying the same locus as the barium stars in the $(P-e)$ diagram, and with a mass function compatible with a WD companion, 
 33\% (=4/12) show a chemical signature of mass transfer in the form of  $s$-process overabundances (from rather moderate -- about 0.3~dex --  to more extreme -- about 1~dex). The only strong barium star in our sample is found in the cluster with the lowest metallicity in the sample (i.e. star 173 in NGC 2420, with [Fe/H] = $-0.26$), whereas the barium stars with mild $s$-process abundance anomalies (from 0.25 to $\sim 0.6$~dex)  are found in the clusters with slightly subsolar metallicities. Our finding
confirms the classical prediction that the $s$-process nucleosynthesis is more efficient at low metallicities, since the $s$-process overabundance is not clearly correlated with the cluster turn-off (TO) mass; such a correlation would instead hint at the importance of the dilution factor.
% or it results from a smaller dilution in the less massive envelope of the giants in clusters with low TO masses. 
We also find a mild barium star in NGC 2335, a cluster with a large TO mass of 4.3~M$_\odot$, which implies that asymptotic giant branch stars that massive still operate the $s$-process and the third dredge-up.}{} 
% 5 {} token are mandatory


   \maketitle
%
%________________________________________________________________

\section{Introduction}

It is difficult to match observational constraints with scenarios of mass transfer involving red giant stars.
Most notably the observed distribution of post-mass-transfer systems in the period -- eccentricity ($P - e$) diagram
does not match the model predictions because models basically predict a bimodal distribution: on the one hand, circular short-period systems ($P < 800$~d),  resulting from Roche-lobe overflow (RLOF),  and on the other hand, eccentric systems with periods longer than about 3000~d that avoided RLOF and always remained detached \citep{1988A&A...205..155B,2003ASPC..303..290P,2008A&A...480..797B,2010A&A...523A..10I}. Canonical models described in these former studies thus predict a gap with no post-mass-transfer systems with periods around 1000~d. However, observations of post-mass-transfer systems like barium stars reveal no such period gap \citep{1998A&A...332..877J}.
To make progress on this issue, we perform in this paper a comparative study of the observed properties  ($P - e$ diagram, mass-ratio distribution) of  samples of pre-mass-transfer and post-mass-transfer systems. 

Barium stars \citep{1951ApJ...114..473B} are a prototypical family of post mass-transfer binaries involving low- or intermediate-mass stars 
\citep[e.g.][]{1980ApJ...238L..35M,1988A&A...205..155B,1990ApJ...352..709M,1998A&A...332..877J}. 
The post-mass-transfer nature of barium stars is made obvious by the chemical anomalies exhibited by the giant primary, for example, strong absorption
lines of ionised barium in its spectrum and of other elements produced
by the $s$ process of nucleosynthesis  \citep[][]{2011RvMP...83..157K}. The $s$-process material has been transferred to the barium star when the former primary,
now a white dwarf (WD), was an asymptotic giant branch (AGB) star.
This binary scenario was convincingly confirmed by the observation
that (almost) all barium stars reside in binary systems 
\citep{1980ApJ...238L..35M,1983ApJ...268..264M,1988A&A...198..187J,1990ApJ...352..709M,1998A&A...332..877J}.
Recently, a radial-velocity monitoring with the HERMES spectrograph \citep{2011A&A...526A..69R} attached to the 1.2 m Mercator telescope from the Katholieke Universiteit Leuven has provided the last remaining orbits (some with orbital periods as long as 50~yr) for a complete sample of barium stars (Jorissen et al. 2016, in preparation).

Here, we make use of this complete set of post-mass-transfer orbital elements  to study the resulting mass function distribution to revisit the mass-ratio distribution of these stars, therefore extending the work of \citet{1990ApJ...352..709M},  \citet{1992btsf.work...26B}, and \citet{1998A&A...332..877J}. For post-mass-transfer systems such as barium stars, the companion should be a CO white dwarf  with a mass larger than 0.51~\Msun\ \citep[for a star of initial mass 0.9~M$_\odot$; e.g. Eq.~66 of ][]{2000MNRAS.315..543H}. This prediction may be tested from the mass-ratio distribution of barium stars.
But before embarking onto this, it is useful to have a comparison sample. In their pioneering work, \citet{1993A&A...271..125B} constructed a sample of normal, field 
G--K giants known to be spectroscopic binaries and found that the mass ratio distribution was most likely close to uniform.
This had the advantage of having field stars similar to the sample of barium stars, allowing for a better comparison. On the other hand, this sample was very heterogeneous, coming from different sources. Since it was not possible to know the masses of these stars, some assumptions had to be made. Since then, we are fortunate enough that a new, homogeneous catalogue has been published. \citet{2007A&A...473..829M} indeed provide spectroscopic orbits for 156 red giants in open clusters, as the final outcome of a very long programme carried out with the CORAVEL instrument \citep{1979VA.....23..279B}. We use this sample as a comparison to our sample of barium-star orbits (and their cooler analogues, the extrinsic S stars).

The paper is constructed as follows. In Sect.~\ref{Sec:methodology}, we present our method to analyse the mass ratio distribution of a sample of binary stars, which we then apply to the sample of \citet{2007A&A...473..829M} in Sect.~\ref{Sec:Merm} and to our enlarged sample of barium and S stars in Sect.~\ref{Sec:Ba}. Section~\ref{Sect:abundances_all} describes an abundance analysis that aims to detect barium stars in open clusters.

\section{Methodology}
\label{Sec:methodology}

For spectroscopic binary systems with only one observable spectrum (SB1), the individual component masses cannot be accessed directly; they instead combine in the spectroscopic mass function $f(m)$:
\begin{equation}\label{Eq:fmm1}
f(m) = \frac{M_2^3}{(M_1+M_2)^2} \sin^3 i = M_1 \frac{q^3}{(1+q)^2} \sin^3 i,
\end{equation}
where $i$ is the (unknown) orbital inclination with respect to the plane of the sky, $M_1$ and $M_2$ are the primary and  secondary masses, respectively, and $q=M_2/M_1$ is the mass ratio.
The mass function (expressed in solar masses M$_\odot$) is derived from  observable quantities
\begin{equation}
f(m) = 1.0385\; 10^{-7} K_1^3 (1 - e^2)^{3/2} P, 
\end{equation}
where $P$ is the orbital period (expressed in days in the above relation), $K_1$ (in km~s$^{-1}$) is the radial-velocity semi-amplitude of the observable primary component,  and $e$  is the eccentricity.

In our case, the primary is the red giant, so if there is a way to know (or assume) its mass, we can, for a given sample, study the distribution of the quantity $Y=f(m)/M_1$. As we can safely assume that the orbital inclination is randomly distributed according to $g(i)=\sin i$, the distribution of $Y$ can thus provide us with the distribution of mass ratios, $d(q)$, that we are looking for \citep[see e.g.][]{1992btsf.work...26B,1993A&A...271..125B,1994InvPr..10..533C,2004RMxAC..21..265P,2010A&A...524A..14B,2012ocpd.conf...41B}.

One way to accomplish this is to look at the distribution of $\log Y$  \citep[and not of $Y$; see][]{2010A&A...524A..14B} and compare this with some given {\it a priori} distributions. Another way is to use a method to numerically invert the equation that links the observed distribution of $Y$ with that of $q$. Here, we use the method 
designed by \citet{1992btsf.work...26B}, which relies on a Richardson-Lucy deconvolution and has proven to be very robust and reliable (see references above).

\section{Binary red giants in open clusters}
\label{Sec:Merm}

 \citet[][ M07 in the following]{2007A&A...473..829M} obtained 
 radial velocities of 1309 red giants in 187 open clusters distributed over the whole sky. These red giants have been monitored with the CORAVEL and CfA spectrometers for 20 years, with a typical accuracy of 0.4 km s$^{-1}$ per observation. They detected 
289 spectroscopic binaries and published orbits for 156 systems with an average of 26 observations per system. The orbital periods range from 41.5 days to 40 yrs and eccentricities from 0 to 0.8. The remaining 133 systems have periods that are too long, an insufficient number of observations, and/or inadequate phase coverage for an orbit determination.

Although the M07 sample of binary red giants in open clusters cannot be considered complete for the reasons listed above, the homogeneity of the data and the
observing strategy nevertheless permit a reliable assessment of the statistical properties of the binary systems of this sample. In particular, one may safely assess that the major incompleteness concerns the systems with the longest orbital periods, above $10^4$~d -- about 30~yr -- , given that the velocity monitoring spanned 20 years.  This selection effect does not bias the intended comparison with the post-mass-transfer sample of barium stars,      which has a similar period cut-off, as we discuss below (Sect.~\ref{Sec:Ba}).
Another advantage of the M07 sample is that it deals with members of open clusters.  We can use the latter to estimate the mass of the red giant needed for Eq.~\ref{Eq:fmm1}. This was carried out in the following way: for each cluster, we used the distance,  reddening, age, and metallicity of the clusters as collected in WEBDA\footnote{\url{http://www.univie.ac.at/webda/}}. The photometric data of the objects are then used to locate them into an H-R diagram. We then used the BaSTI\footnote{\url{http://albione.oa-teramo.inaf.it/}}
isochrones corresponding to the adequate metallicity, and corrected for the given reddening and distance, to determine the mass of the stars. 
We preferred to use this method than simply use the turn-off of the cluster as the mass of the giants, as this should lead to a more precise value and an independent check of  membership of the star. In fact, this methodology was usable on only 124 systems, which is the sample we  use in the following. 
Of the 32 systems for which we could not find an adequate solution, 28 were flagged by M07 as non-cluster members but field stars. The remaining four (NGC 2489 25,
NGC 2925 92,
Ru 79 2, and
Tr 26 201) are supposedly cluster members, but we do not consider them further here, as their small number is not going to degrade  the statistics.

\begin{figure}
\includegraphics[width=9cm]{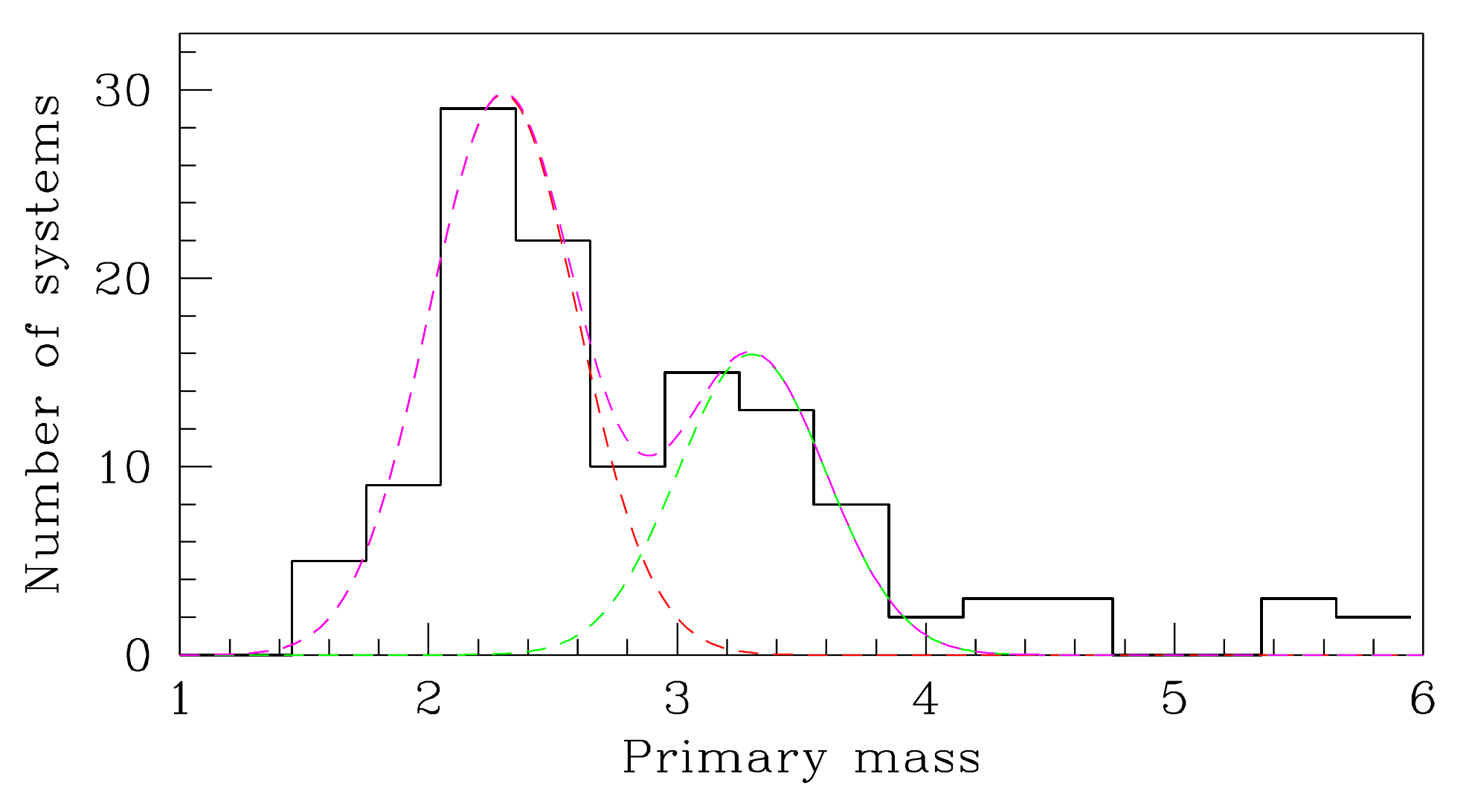}
\caption{\label{Fig:mermass} Distribution of the primary masses, $M_1$, for the M07 sample of red giants is shown as the black histogram.
We also show a Gaussian distribution centred on 2.3~M$_\odot$ and with $\sigma=0.3$~M$_\odot$  with the red dashed line, while the thin dashed  green line is a Gaussian centred on 3.3~M$_\odot$ and with $\sigma=0.3$~M$_\odot$.}
\end{figure}

The resulting mass distribution is shown in Fig.~\ref{Fig:mermass} and Table~\ref{Tab:cluster_mass}. Typical errors on the mass
we so derive are between 0.05 M$_\odot$ and 0.2 M$_\odot$, i.e. between 1\% and 9\%. Such a small error has no consequence on the mass ratio distributions we derive, given the weak dependence of the mass ratio on the primary mass, for a given spectroscopic mass function (see Eq.~\ref{Eq:fmm1}).
The bulk of the primary-mass distribution we obtain can be approximated by a Gaussian distribution centred on 2.3~M$_\odot$ and with a standard deviation $\sigma=0.3$~M$_\odot$, although there are some additional systems with masses between 3 and 4~M$_\odot$ (with a secondary peak at 3.3~M$_\odot$). The mean mass of the red giants in this sample is 2.9~M$_\odot$. In the following, we carry out the analysis using either the actual primary-mass distribution as determined through isochrone fitting or  using a single  Gaussian distribution as first approximation.

%\begin{figure}
%\includegraphics[width=9cm]{mermelogP}
%\caption{\label{Fig:uvpti}
%}
%\end{figure}

\begin{figure}
\includegraphics[width=9cm]{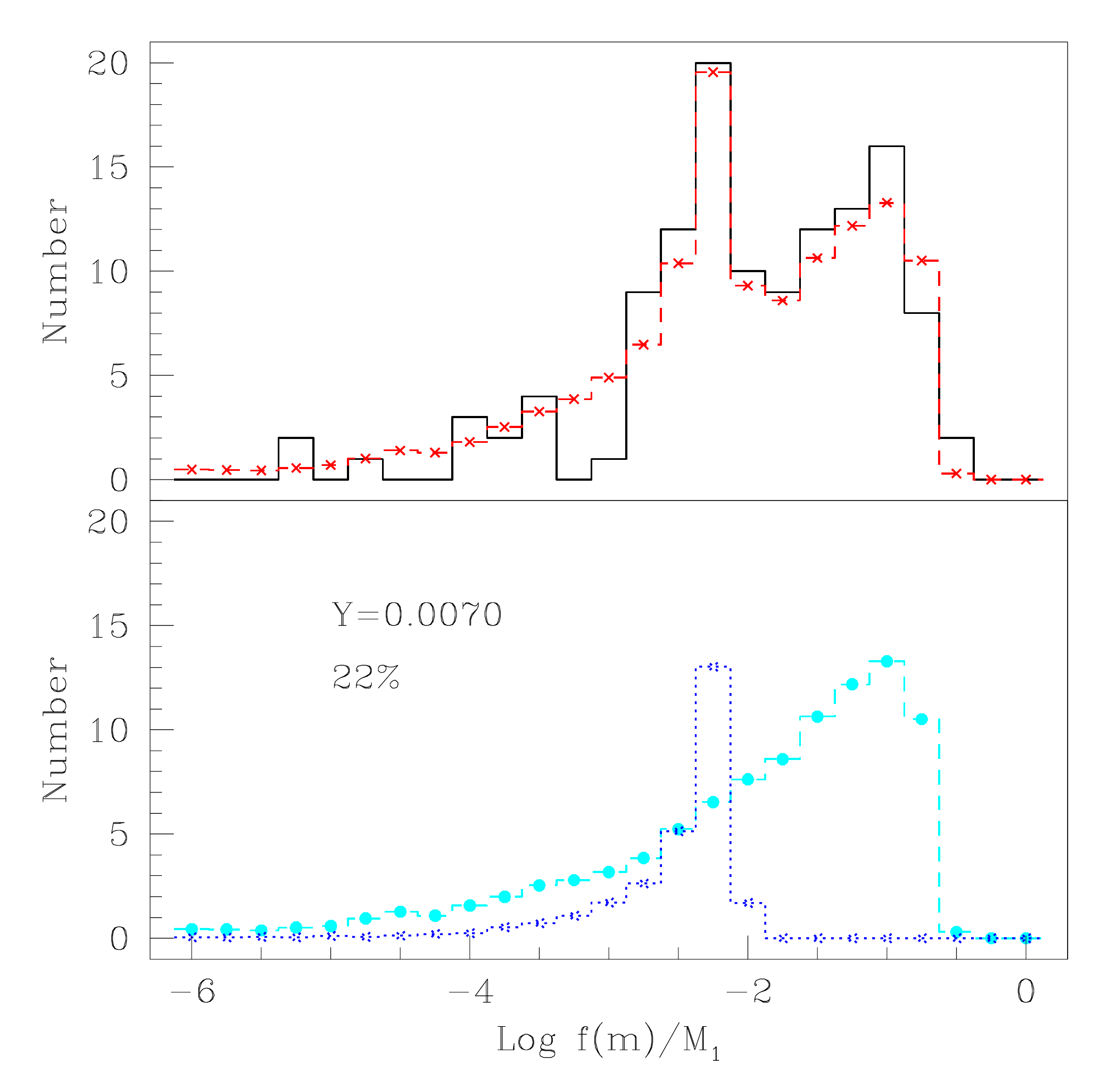}
\caption{\label{Fig:mermfmall} (Upper panel) Black histogram shows the distribution of the logarithm of the reduced mass function, $Y=f(m)/M_1$, for the sample of spectroscopic binaries from M07, while the red dotted line connecting filled dots is our best model (see text). (Lower panel) 
This shows the two components entering into our best-fit model using a Monte Carlo approach:
The cyan dashed line connecting solid dots is the distribution that one would expect for a uniform mass-ratio distribution, while the blue dotted line connecting crosses corresponds to having 22\% of systems with a Gaussian distribution of $Y$, centred around $Y=0.007$ and a standard deviation of 0.0009. The best-fit red line in the upper panel is the sum of these two distributions.  %The red line is therefore coincident with the cyan line for large values of $f(m)$.
}
\end{figure}

The resulting distribution of $\log Y$ for the M07 sample, using the primary masses we determined, is shown in Fig.~\ref{Fig:mermfmall}, where we also show the distribution (cyan dashed line in the bottom panel) we would expect if we had a uniform mass-ratio distribution (MRD), as was found by 
\citet{1993A&A...271..125B} for their sample of red giant spectroscopic binaries. Clearly, the observed distribution has an over density of systems
with $-3 < \log Y < -2$ compared to the uniform distribution.  As an illustration, a value of 
$\log Y = -2$ corresponds to a mass ratio of 0.3, when taking the mean value of $\sin^3 i$. For a primary mass of 2.3~M$_\odot$, such a mass ratio corresponds to a 
secondary mass of 0.69~M$_\odot$. Thus, it appears that the observed distribution of $\log Y$ indicates that there are more systems
with such a mass ratio (or corresponding secondary mass) than one would expect from a uniform distribution. There is no reason to expect such an excess of 0.6--0.7~M$_\odot$ stars, if they were on the main sequence, unless some process during star formation led to a peak at this mass ratio. This is, however, neither seen in the analysis of field red giant stars, nor
in the study of solar-like stars (Halbwachs et al. 2003). Such a mass does correspond, however,  to the typical mass of CO WDs \citep{2010ApJ...712..585F,2013ApJS..204....5K}, and one could thus assume that there is a fraction of systems in the M07 sample that are post-mass-transfer systems, i.e. systems in which the  (present) red giant was initially the least massive of the two stars and the primary already evolved to the WD stage. Such systems are similar to barium and S stars, although they may not all be contaminated in $s$ process (this is checked in Sect.~\ref{Sect:abundances}), and would be the possible descendants of blue straggler stars seen in open clusters. Such post-mass-transfer giants should thus be somewhat bluer (about 0.05 - 0.07~mag in $B-V$ for a mass difference of 0.2~\Msun) than the other giants. This effect was not seen when the giant masses were derived from isochrone fitting (see earlier in this Sect.), which is probably because the available photometric data was not accurate enough.

We have therefore tried to add, on top of the uniform mass-ratio distribution, a sample of post-mass-transfer systems, represented by systems with a peaked Gaussian distribution of 
$<~Y~>~=~Y/\sin ^3 i$, as was found for barium stars \citep[][ see also Sect.~\ref{Sec:Ba}]{1988covp.conf..403W}. Such a peaked distribution would mean that the mass ratio is peaked, as expected for a WD mass distribution. We have therefore searched for the best fit, by means of $\chi ^2$ minimisation to the observed distribution, in which we add three free parameters: the fraction of systems belonging to the post-mass-transfer population, $n_p$, and the mean and standard deviation of the Gaussian distribution, $Y_p$ and $\sigma_p$. The result of the best fit, corresponding to a reduced $\chi^2_r=1.035$, is shown in Fig.~\ref{Fig:mermfmall}, and is given by $n_p = 0.22 \pm 0.02$, $Y_p=0.0070$, and $\sigma_p=0.0009\pm 0.0001$. Such a value of $Y_p$ would translate into a peaked distribution of mass ratios around $q=0.216$, which, assuming the  mean mass of red giants in the sample, 2.9~M$_\odot$, indicates a peaked secondary mass around 0.63~M$_\odot$, exactly as expected for WDs. For primary masses of 2.3 and 3.3~M$_\odot$, as suggested from Fig.~\ref{Fig:mermass}, the secondary mass amounts to 0.50 and  0.71~M$_\odot$, respectively. 
We conclude that we need to add $\sim$22\% of post-mass-transfer systems containing a WD to reproduce the observed distribution of $\log Y$ for giants in open clusters.

\begin{figure}
\includegraphics[width=9cm]{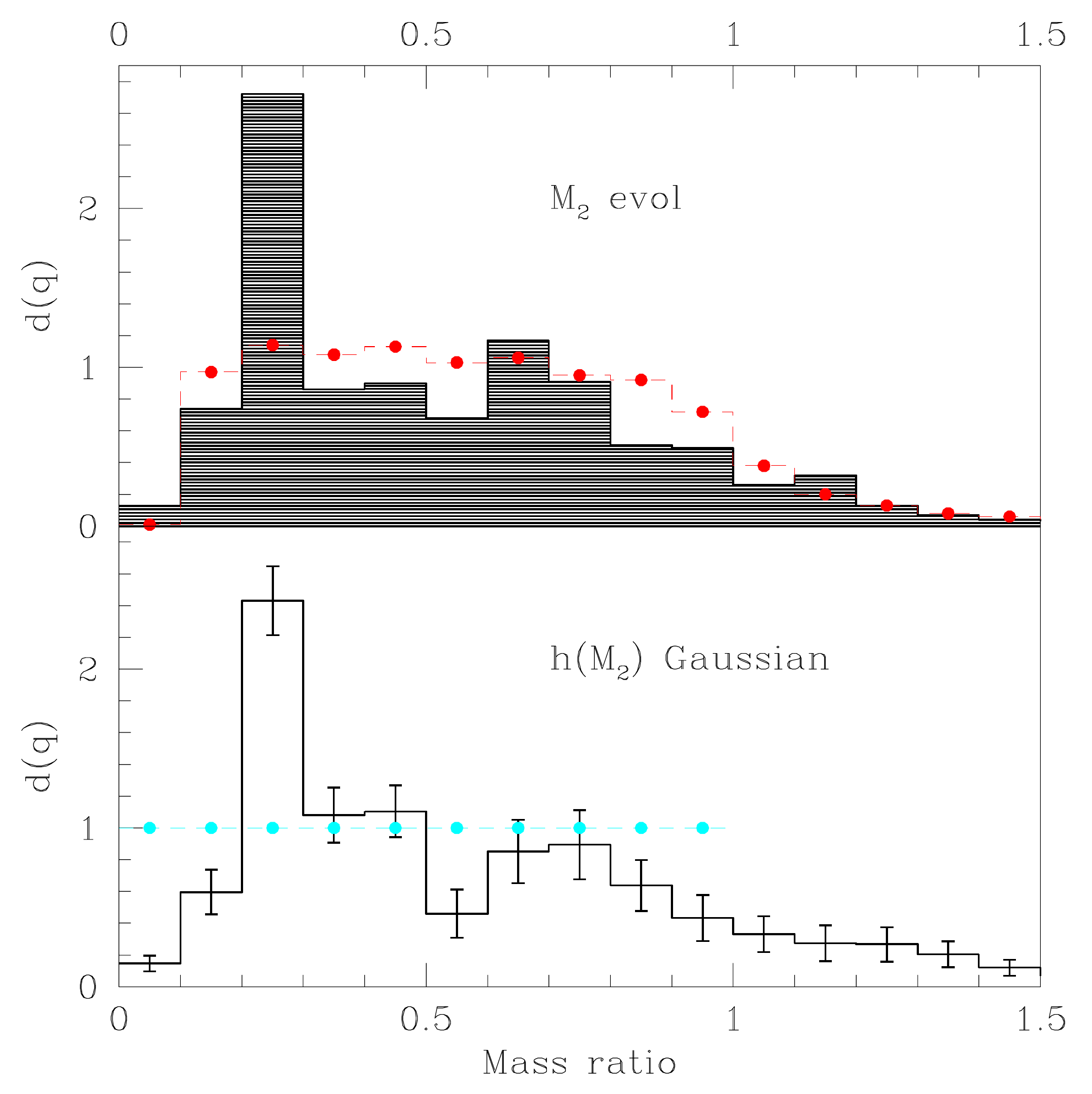}
\caption{\label{Fig:mermfq2gau} Mass ratio distribution corresponding to the M07 sample as derived with the Richardson-Lucy algorithm, when using the primary mass derived from the fit of the isochrones (top panel) or assuming a Gaussian distribution (lower panel). For the latter, we have run 1,000 simulations and indicate the  $1~\sigma$ error bars associated with these. In the top panel, we also show, with the red dashed line connecting heavy dots, the distribution we obtain with our inversion method when giving as input a uniform distribution (see text), while in the lower panel the dashed cyan line connecting dots just shows a uniform distribution of mass ratios for comparison.
}
\end{figure}

Of course, instead of assuming some functional form for the MRD and although we seem to be obtaining a very good fit to the observations, it may be more appropriate to use an inversion technique to obtain the MRD directly from the observed distribution of $Y$, without any a priori assumption on the form of the MRD. As mentioned above, this can be carried out with the Richardson-Lucy deconvolution method \citep{1974AJ.....79..745L}. We have used this method and the results are shown in Fig.~\ref{Fig:mermfq2gau}. In the upper panel, we show the MRD obtained when using the primary masses we derived from isochrones. This figure shows what is now a familiar result: the MRD is generally very uniform (with a small scatter due to the small number of systems) but with an additional strong peak of systems with a mass ratio between 0.2 and 0.3. For the mean primary mass of this sample, such a mass ratio corresponds to secondary masses between 0.58 and 0.87~M$_\odot$, i.e. typical masses of WDs, in agreement with what we found above. We also note that our computed distribution shows a few systems with a mass ratio above 1. As the M07 sample is composed of single-lined binaries with a red giant primary, and since the most massive star should have evolved first, these few systems  with $q>1$ are most probably an artefact of the method (due to the limited resolution) and, in some cases, of an underestimate of the primary mass. Similarly, the quasi-absence of systems with mass ratios in the first bin $0<q<0.1$ is most likely due to an observational bias, as systems with very low mass ratios have  radial-velocity amplitudes that are too small (and thus too small $f(m)$) to be detected. To illustrate this, we show in the top panel of Fig.~\ref{Fig:mermfq2gau} the result of an experiment we conducted. We created an artificial sample of systems drawn from a uniform distribution of mass ratios, and associated to each of them an inclination, assuming $g(i) = \sin i$. We then added a 10\% error to the so-derived spectroscopic mass function and removed all systems that had a $f(m)$ smaller than the smallest in our sample. We then applied our inversion method to derive the mass ratio distribution and show this as the red dashed curve. We can see that we account for the lack of systems in the first bin and the systems with apparent mass ratio above one.

The lower panel of the same figure shows the MRD we obtain if, instead of using the primary masses we derive from isochrone fitting, we assume that the primary-mass distribution is a Gaussian centred on 2.3~M$_\odot$ with a standard deviation of 0.3~M$_\odot$. We have run 1,000 simulations, generating  the primary mass of a given system according to this distribution every time. This allows us to estimate the typical error bar on each bin in the histogram. As one can see, there is practically no difference resulting from the use of the actual primary-mass distribution or from this Gaussian first approximation. We use this fact later when studying barium and S stars.

While this work was well underway, we came across the paper by \citet{2014psce.conf...63N}, which also analysed the sample of cluster giants of Mermilliod et al. as part of a study of A stars. The analysis in this study is, however, restricted to assuming a flat mass ratio distribution for the companions of the red giants, and comparing the distribution of the logarithm of $f(m)$ so obtained, assuming random inclination. North then found, as we did, that the peak in this distribution requires an additional component, which he models as a distribution of WD companions following a Gaussian centred around 0.6~M$_\odot$ and with a dispersion of 0.03~M$_\odot$. To reproduce the peak, \citet{2014psce.conf...63N} needs a relative number of WD companions of about 23\%. His results are thus in agreement with ours, but our method appears more rigorous, with a minimisation method used to find the final set of parameters. Moreover, our inversion technique provides us with a direct mean to obtain the true mass ratio distribution. %and thus confirm our result. 
Finally, North argues that the proportion of 23\% of WD companions can be easily accounted for by assuming a Salpeter initial mass function for the original primaries (the WD progenitors thus) and a uniform (or quasi-uniform) mass-ratio distribution, given the narrow age and mass distributions of the observed giants in the clusters. 

\subsection{Looking for the post-mass-transfer objects}
\label{Sect:post-mass-transfer}

Our results above have shown that the red giant binaries in open clusters contain a fraction, 22\%, of systems with a WD companion. As this results from a statistical analysis, it is of course not possible to identify which are these post-mass-transfer systems. One possibility would be to make use of a well-known property of stars in the period -- eccentricity ($P - e$) diagram, as proposed in Jorissen \& Boffin (1992) and Boffin et al. (1993). It is known that most binary samples lack systems with small eccentricities at long periods. Since this gap is already present among pre-main-sequence binaries \citep{1992btsf.work..155M}, it must be a signature imprinted by the binary formation processes; more precisely, no binary systems form in circular orbits. Subsequent circularisation of the binary system results either from tidal effects (for the shortest systems) or from mass-transfer processes. As a result, the exact extent of this gap, which we call the long-period gap, depends upon the kinds of systems under consideration. In FG main-sequence binaries \citep{1991A&A...248..485D} for example, no systems with $e<0.1$ and $P> 130$~days are found. But for post-mass-transfer systems such as barium stars,  the gap is found at $e<0.05$ and $P>1000$~d (see the top panel of Fig.~\ref{Fig:elogpmerm}). This narrowing of the gap must clearly be attributed to the mass transfer that occurred in those systems.
 Therefore, it was hypothesised by Boffin et al. (1993) that the systems with red giants that are in the area $e<0.1$ and $P> 130$~d could be post-mass-transfer systems. We test a generalised version of this assumption here using two different methods: 
\begin{itemize}
\item We look at the $f(m)$ distribution for this subsample and see whether it reveals possibly a larger percentage of WD companions in the sample;
\item We perform a chemical abundance analysis for a few stars located in this region with the aim to identify whether they are $s$-process enriched like barium stars.
\end{itemize}

In addition, we establish the mass-ratio distribution for a subsample of short-period systems and see whether, as expected, the fraction of post-mass-transfer systems has been considerably reduced.

\subsubsection{The mass function}
\begin{figure}
\includegraphics[width=9cm]{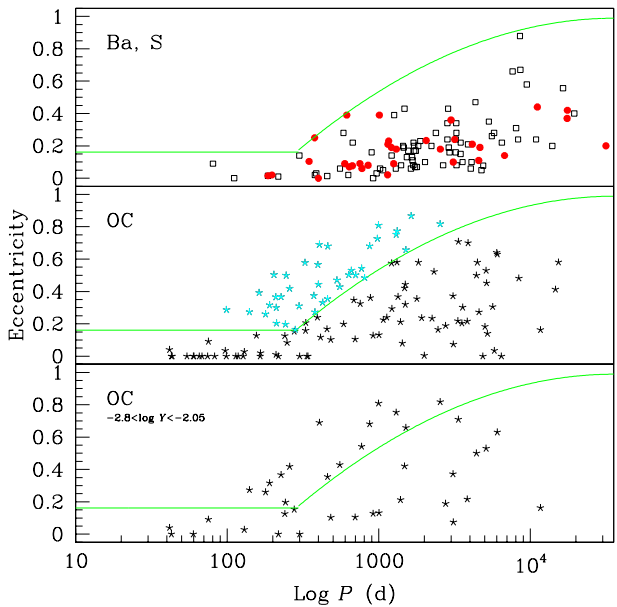}
\caption{\label{Fig:elogpmerm} 
Orbital period -- eccentricity diagrams for the various samples considered here. The upper panel shows orbits of barium (open squares) and S stars (red dots), as well as a conservative envelope that encompasses all the points (green line). The middle panel shows red giants in binary systems from M07, separated according to the envelope defined above (black and cyan symbols). The lower panel shows only those systems from M07 with mass functions in the range  $-2.8 < \log Y < -2.05$, that is, those that are more likely to be post-mass-transfer systems  (however, about half of these systems are not, according to the bottom panel of Fig.~2).
}
\end{figure}

\begin{figure}
\includegraphics[width=9cm]{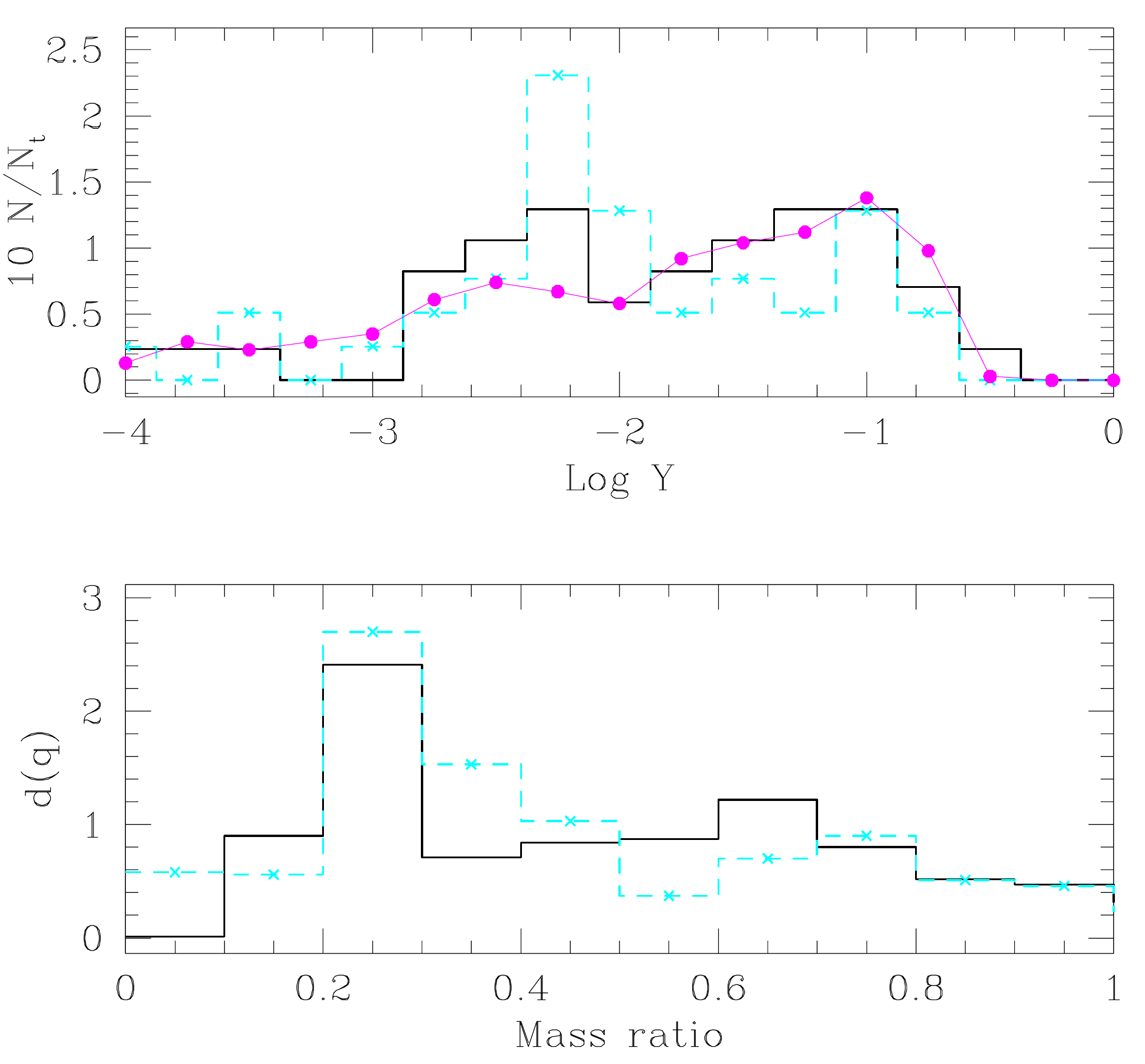}
\caption{\label{Fig:pmt} Distribution of $\log Y$ (top panel) and mass-ratio distribution (bottom) for the two subsamples of M07, according to their position in the  $P - e$ diagram; the heavy black curve corresponds to those systems below the envelope shown in Fig.~\ref{Fig:elogpmerm}, while the cyan dashed line corresponds to those above the envelope. The latter should therefore not contain post-mass-transfer systems, i.e. systems with a WD companion. In the top panel, the magenta line connected by heavy dots represents the distribution expected in case of a uniform distribution. It can be seen that both subsamples deviate from the uniform distribution and the mass ratio distributions are barely distinguishable, even if the systems with  large eccentricities seem to be more peaked around $\log Y = -2.3$.}
\end{figure}

%{\it Mention here that we cannot distinguish systems in e-log P diag for those that have $-2.8 < \log Y < -2.05$, i.e. those contributing to the peak of WDs...}
Figure~\ref{Fig:elogpmerm} shows the ($P - e$) diagram for our sample of barium and S stars (to be described in  Sect.~\ref{Sec:Ba}) and open cluster stars. As can be seen in the top panel, the eccentricity of barium and S stars is delineated by an envelope, with the mean eccentricity at a given orbital period being smaller for barium and S stars than for normal giants (Boffin et al. 1993) or open cluster giants.
To be conservative, we defined this envelope as the curve shown in Fig.~\ref{Fig:elogpmerm} and defined by
\begin{eqnarray}
\label{Eq:ethresh}
e = 0.16 && \mathrm{if} \log P < 2.47 \\
e = -2.90 +1.71 \log P - 0.19 (\log P)^2 && \mathrm{if} \log P \geq 2.47.
\end{eqnarray}

We then separated the sample of M07 giants into two subsamples, depending on which side of the envelope the giants are located, with the null hypothesis being that the subsample below the envelope should contain more post-mass-transfer systems, i.e. more systems with a WD companion, while the subsample above the envelope should not contain any. For these two subsamples, we applied the Richardson-Lucy inversion technique to derive the respective mass-ratio distributions as shown in the lower panel of Fig.~\ref{Fig:pmt}. As can be seen, there is basically no difference between the two subsamples,  as both show an excess of low-mass companions. 

We also tried the opposite approach. We selected all systems with $-2.8 < \log Y < -2.05$ from the M07 sample, i.e. those systems that appear over-represented in Fig.~2 compared to a uniform distribution and which therefore correspond to the white dwarf population. These systems are shown in the $P - e$ diagram in the lower panel of Fig.~\ref{Fig:elogpmerm}, where it can be seen that they are found on both sides of the post-mass-transfer system's envelope.   It should be remembered, however, that the post-mass-transfer systems represent only about half of the systems in the  $-2.8 < \log Y < -2.05$ range (more precisely 21/37 = 57\%), as may be judged from Fig.~2. 

Thus it appears that it is not at all easy to separate  those systems that contain a WD from those that do not in the $P - e$ diagram. 

%{\it Derive mass ratio distribution for these systems in e-log P region}

\subsubsection{Chemical enrichment}

We performed an abundance analysis of the giants with long periods and low eccentricities to identify those possibly bearing the chemical signature of mass transfer from a thermally pulsing AGB companion, in the form of overabundances of $s$-process elements. In other words, we have looked for possible barium stars among the M07 binaries. Although this abundance study is deferred to Sect.~\ref{Sect:abundances} below, 
we present its conclusions here. One barium star with strong anomalies is confirmed in NGC 2420 (star 173, with [Fe/H] = -0.26), and three more with mild anomalies (of the order of 0.3 to 0.5~dex) are found 
in solar-metallicity clusters (Table~\ref{Tab:final_abundances}). Among the  12 stars studied [all located in the same region of the $(P - e)$ diagram as barium stars], 4 (or  33\%) thus exhibit some abundance anomalies, a fraction well in line with the expectation from the mass-function analysis.

\subsection{The short-period systems}
\begin{figure}
\includegraphics[width=9cm]{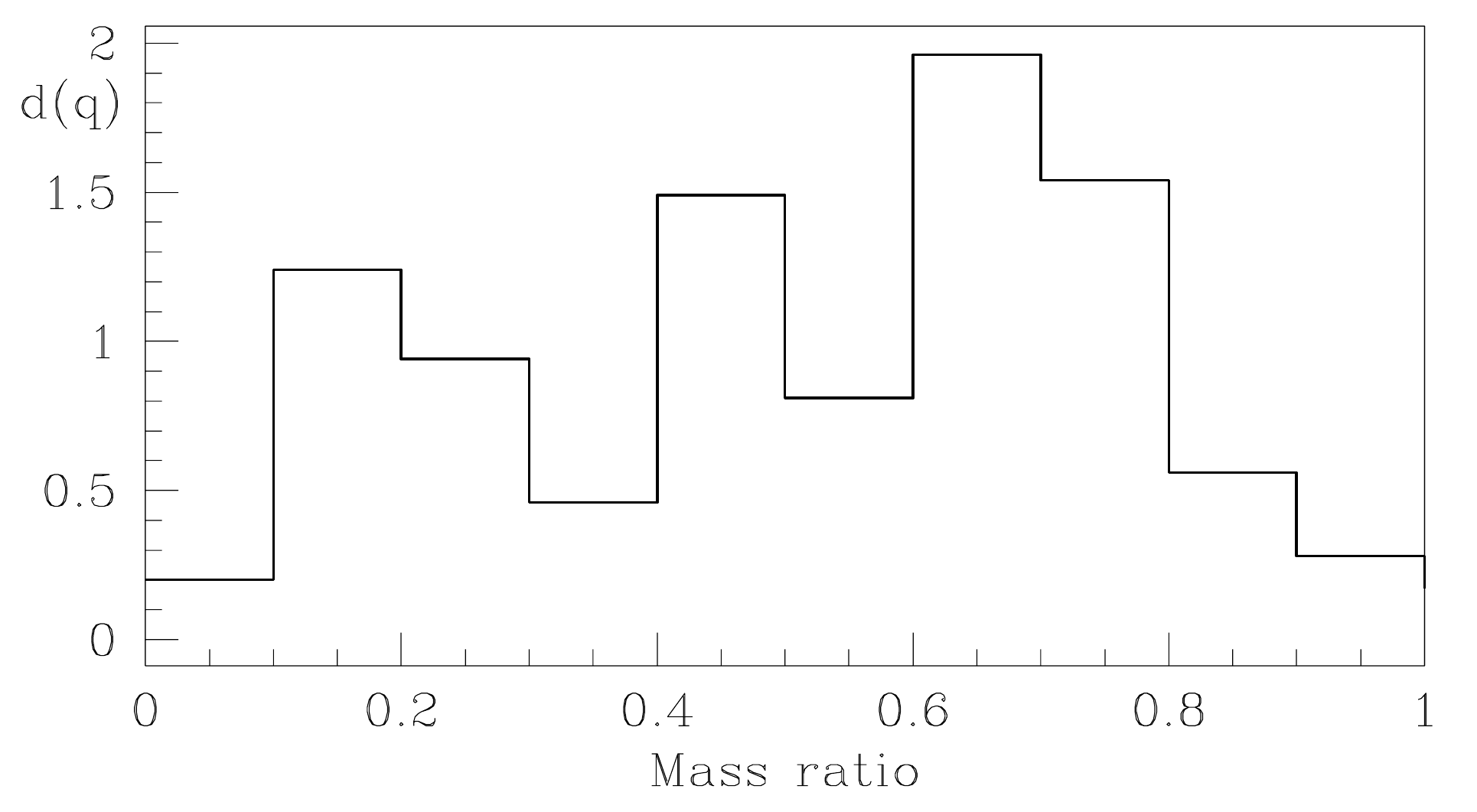}
\caption{\label{Fig:mermfqshortP}
Mass-ratio distribution corresponding to all the systems from the M07 sample with an orbital period smaller than 180 d, as derived from the Richardson-Lucy algorithm, when using the primary masses derived from the isochrone fits.
}
\end{figure}

Finally, as post-mass-transfer systems should have orbital periods large enough to have avoided the common envelope phase, we should not expect many of these in the M07 sample at small periods. We analysed the mass ratio distribution as obtained with the Richardson-Lucy method for a subsample of M07, taking into account all systems with an orbital period below 180 days, where this value is somewhat arbitrary. In our sample of barium and S stars (see below), we have only two systems with periods below this value.
The result is shown in Fig.~\ref{Fig:mermfqshortP}, which, given the small sample (24 systems), is compatible with a uniform distribution of mass ratios, without the need to add any WD companion population.

 \section{Barium and S stars}
        \label{Sec:Ba}
        
        \begin{figure}
\includegraphics[width=9cm]{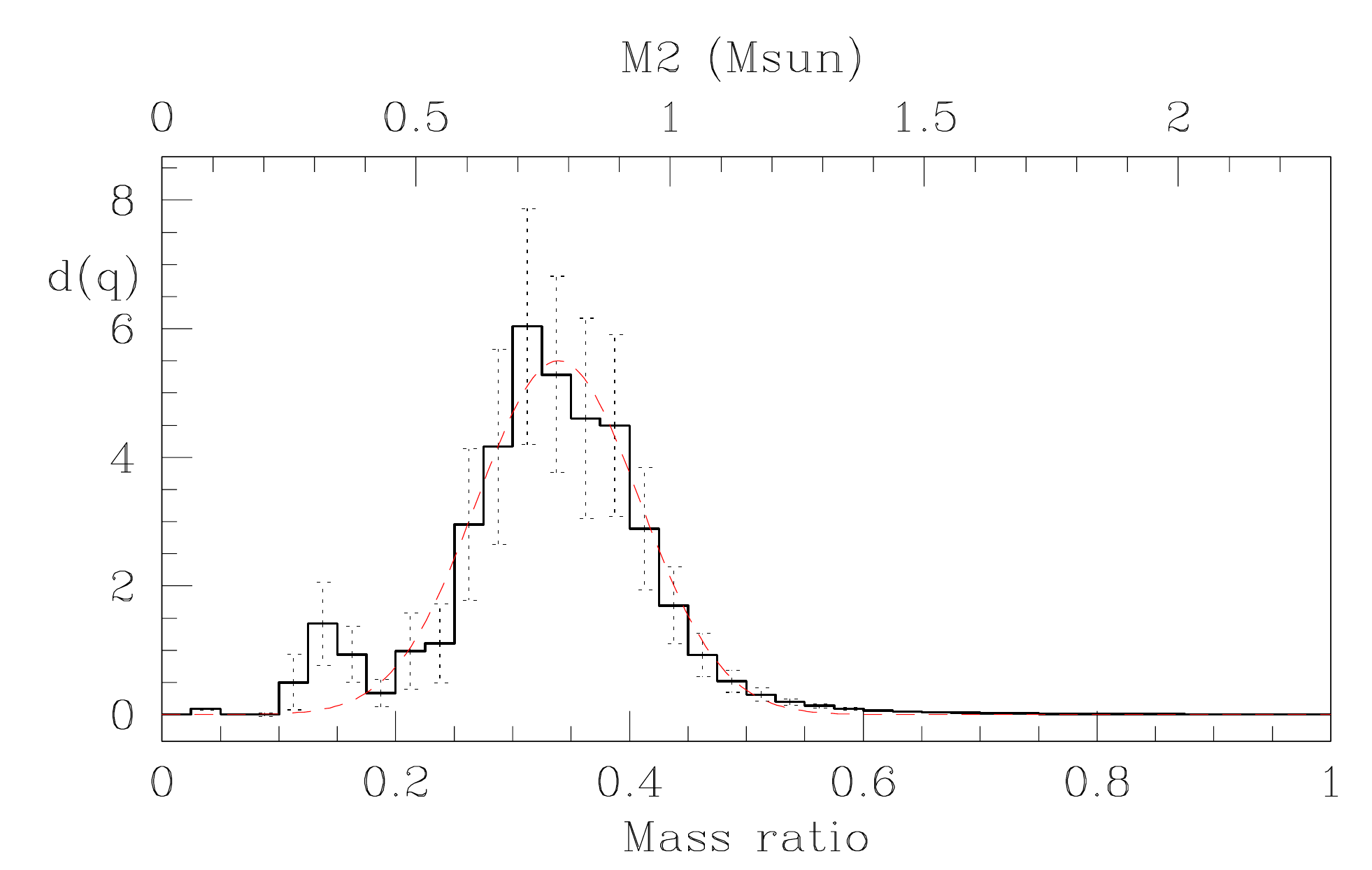}
\caption{\label{Fig:bafq} The mass-ratio distribution for barium stars as derived from the Lucy-Richardson inversion method. A Gaussian fit centred on 
$q= 0.33$ and with $\sigma= 0.065$ is shown for comparison (dashed red curve). The upper horizontal scale gives the WD mass, adopting 2.3~\Msun\ for the average mass of the giant star and  0.3~\Msun\ for its standard deviation. The error bars are 1$\sigma$ errors, based on 1000 Monte Carlo simulations.
}
\end{figure}

%\begin{figure}
%\includegraphics[width=9cm]{bafq2and3forM1}
%\caption{\label{Fig:bafq2} Same as Fig.~\ref{Fig:bafq} but ...
%}
%\end{figure}

\begin{figure}
\includegraphics[width=9cm]{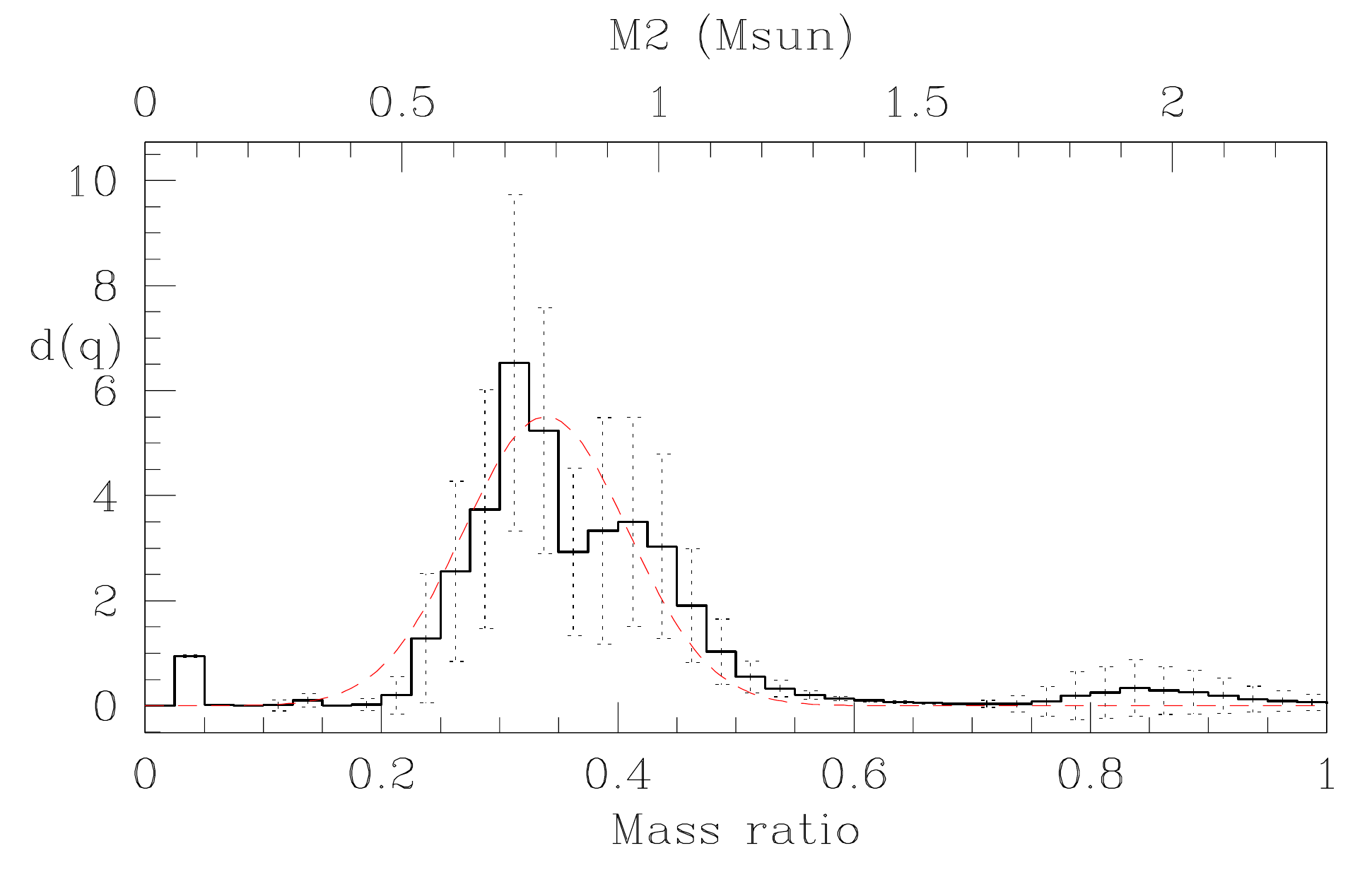}
\caption{\label{Fig:sfq} Same as Fig.~\ref{Fig:bafq} for S stars. 
%Here the Gaussian is centred on $q=0.36$.
}
\end{figure}

\begin{figure}
\includegraphics[width=9cm]{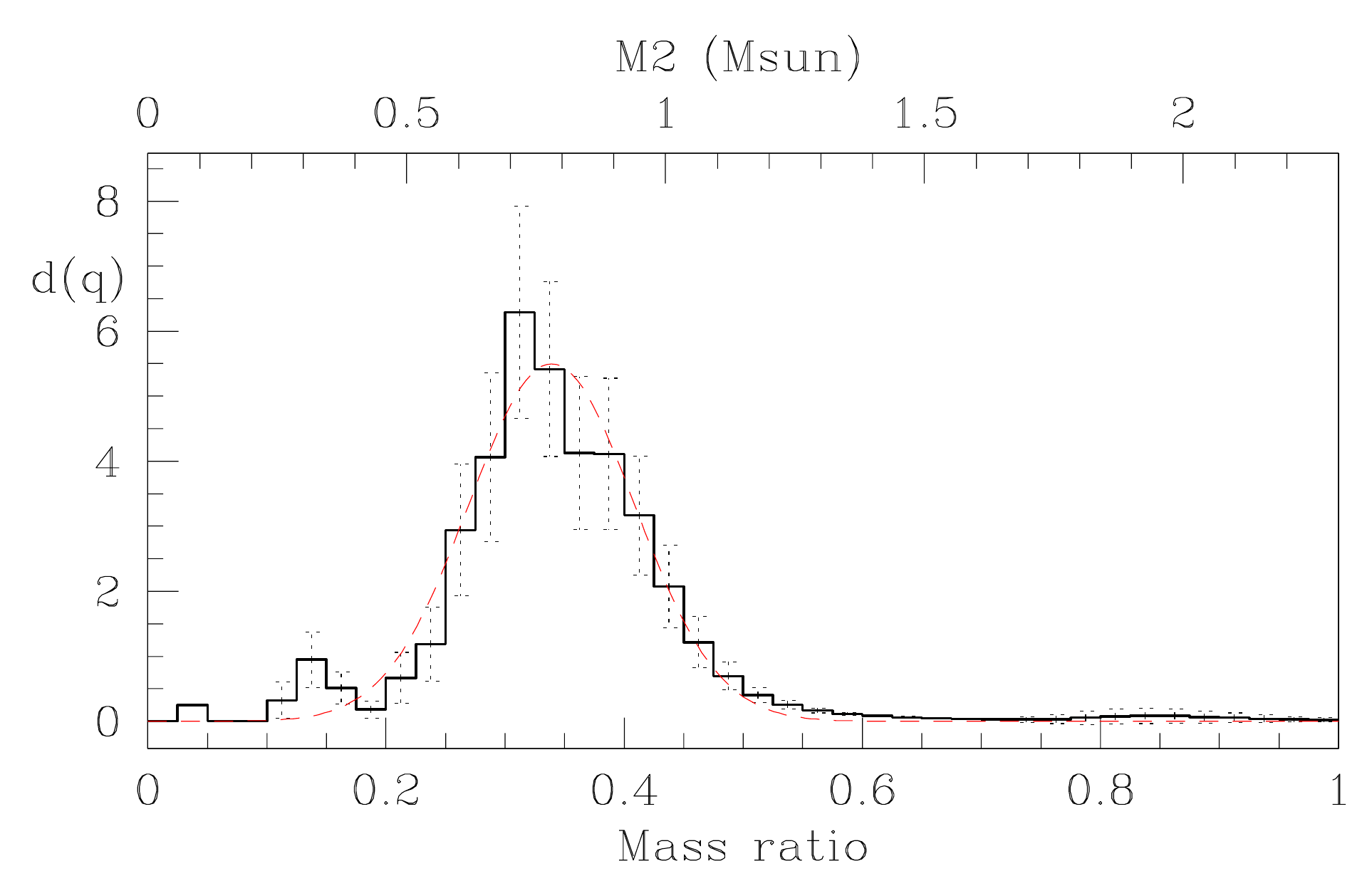}
\caption{\label{Fig:basfq} Same as Fig.~\ref{Fig:bafq} for the combined sample of barium and S stars. 
%A Gaussian fit centred on  $q= 0.34$ and with $\sigma= 0.065$ is shown for comparison.
}
\end{figure}

A companion paper (Jorissen et al. 2016, in preparation) complements earlier studies collecting orbital elements for barium and S stars \citep{1980ApJ...238L..35M,1983ApJ...268..264M,1988A&A...198..187J,1990ApJ...352..709M,1998A&A...332..877J},
and finally provides the longest orbits among barium and S stars, some with orbital periods up to 50 years. These orbits are obtained in the framework  of the ongoing HERMES/Mercator radial-velocity monitoring \citep{2010MmSAI..81.1022V,2013EAS....64..163G}; see Jorissen et al. (2016, in preparation) for detailed information. For the sake of completeness, the full list of mass functions, periods, and eccentricities currently available for barium and S stars is given in Tables~\ref{Tab:strongBa} -- \ref{Tab:S}. 

In the remainder of this paper (especially Sects.~\ref{Sect:MRD_Ba} and \ref{Sect:abundances}), it will sometimes be necessary to distinguish between the so-called mild and strong barium stars. This distinction is made on the "Ba index" introduced by \citet{1965MNRAS.129..263W}, and reflecting the strength  of the Ba lines, based on visual inspection, on a scale from Ba1 to Ba5, Ba5 corresponding to the strongest lines. In this and our past studies, we associate Ba1 - Ba2 indices with mild barium  stars and Ba3 - Ba5 indices with strong barium stars.

\subsection{Mass-ratio distribution}
\label{Sect:MRD_Ba}

The derivation of the MRD of barium stars is plagued by the uncertainty existing on their masses. Unlike giants in clusters, there is no direct way to assess the masses of field giants. \citet{1997A&A...326..722M}  used a Bayesian method to infer barium-star masses, based on their location in the Hertzsprung-Russell diagram, using Hipparcos parallaxes. They conclude that mild and strong barium stars have somewhat different mass distributions, as mild and strong barium stars are characterised by masses in the range 2.5~--~4.5~\Msun\  and 1~--~3~\Msun, respectively. Since the MRD analysis treats mild and strong barium stars together, a distribution that is intermediate between these two seems appropriate. We therefore adopt a Gaussian distribution centred on 2.3~M$_\odot$ and with a standard deviation  
of 0.3~M$_\odot$ for the primary mass of the barium systems, similar to the M07 sample. Fortunately, as shown with our study of the M07 sample, using the real primary-mass distribution or a simplified version of it (Gaussian distribution) does not fundamentally alter the results.

We derived the MRD of barium stars from the 72 orbits listed in Tables~\ref{Tab:strongBa} -- \ref{Tab:mildBa}, and adopted the Gaussian mass distribution described above.
The outcome of this procedure is shown in Fig.~\ref{Fig:bafq}, where it is apparent that the distribution is very peaked around $q \sim 0.30$. A Gaussian distribution that best fits the results  is centred on $q = 0.33$ with a standard deviation of 0.065. This result is very robust with respect to the choice of  the parameters of the barium-star mass distribution ($M_1$). The average $q$ varies from 0.3 for $<M_1> = 3$~M$_\odot$ to 0.4 for $<M_1> = 1.5$~M$_\odot$, whereas $\sigma(q)$ stays at 0.065. This mass ratio then corresponds to companion masses $M_2$~=~0.60, 0.76, and 0.90~M$_\odot$ for $M_1 = 1.5$, 2.3, and 3~M$_\odot$, respectively.
The dispersion around these values for the companion mass $M_2$ cannot be derived with certainty, since it depends upon the adopted dispersion around $M_1$. The value $\sigma(q) = 0.065$ derived from the observed mass-function distribution implies $\sigma(M_2) = 0.03$~M$_\odot$ if  $\sigma(M_1) = 0.30$~M$_\odot$ (and  $M_1 = 2.3$~M$_\odot$), or $\sigma(M_2) = 0.014$~M$_\odot$ if  $\sigma(M_1) = 0.34$~M$_\odot$.

In any case, the above masses for the companion are consistent with carbon-oxygen WDs. 
This is no surprise \citep{1990ApJ...352..709M,1998A&A...332..877J,2000IAUS..177..269N} as barium stars are now well established to be post-mass-transfer systems with WD secondaries \citep{1990ApJ...352..709M,1992btsf.work..110J,2016A&A...586A.151M}. 
The above result  was expressed in a slightly different way by McClure \& Woodsworth (1990) who stated that for barium systems, the value of $Q = M_2^3 / (M_1 + M_2)^2$ can be considered a constant, $Q=0.046$~\Msun, as confirmed in Sect.~\ref{Sect:sigma}. 
 To reconcile the mean value of the WD mass $M_2$ in barium systems with that of field DA WDs  \citep[$0.647\pm0.014$~M$_\odot$, from gravitational redshifts;][]{2010ApJ...712..585F},
a typical barium-star mass of 1.62($\pm0.20)$~M$_\odot$ needs to be adopted.
 We can only hope to determine precisely the mass of the barium stars with Gaia and its expected delivery of many astrometric binaries. Only then will we be able to check whether or not there is a significant difference in the mass distribution of WDs in barium-star systems and in the field, and thereby provide some constraints on the mass-transfer mechanism. 

\begin{table}[htbp]
 \caption{\label{Tab:Gau} Best-fit parameters for the Gaussian distributions of primary and secondary masses, and $Q \equiv   M^3_{2}/(M_{1}+M_{2})^2$, where all masses are expressed in M$_\odot$.  Only $Q$ is constrained by the fit, not $M_{1}$ and $M_{2}$ individually; therefore the first line for each category (Ba or S) lists the pairs that were used in the fit, and the following  lines list other possible combinations of  $M_{1}$ and $M_{2}$ yielding the same $Q$.
}
      \begin{tabular}{lccccc} % Column formatting, @{} suppresses leading/trailing space
     \hline
      Sample    &$M_{1}$ & $\sigma_{1}$ & $M_{2}$ & $\sigma_{2}$ & $Q$\\
    \hline

Ba  & 1.6 & 0.4 & 0.58 & 0.04 & 0.041\\
      & 1.46 & & 0.55 & & 0.041\\
      & 2.04 & & 0.67 & & 0.041\\
      & 2.46 & & 0.75 & & 0.041
\medskip\\
S   &  1.5 & 0.4 & 0.58 & 0.04 & 0.045\\
     & 1.60 & & 0.60 & & 0.045\\
    \hline
   \end{tabular}
\end{table}

\begin{figure}
\includegraphics[width=9cm]{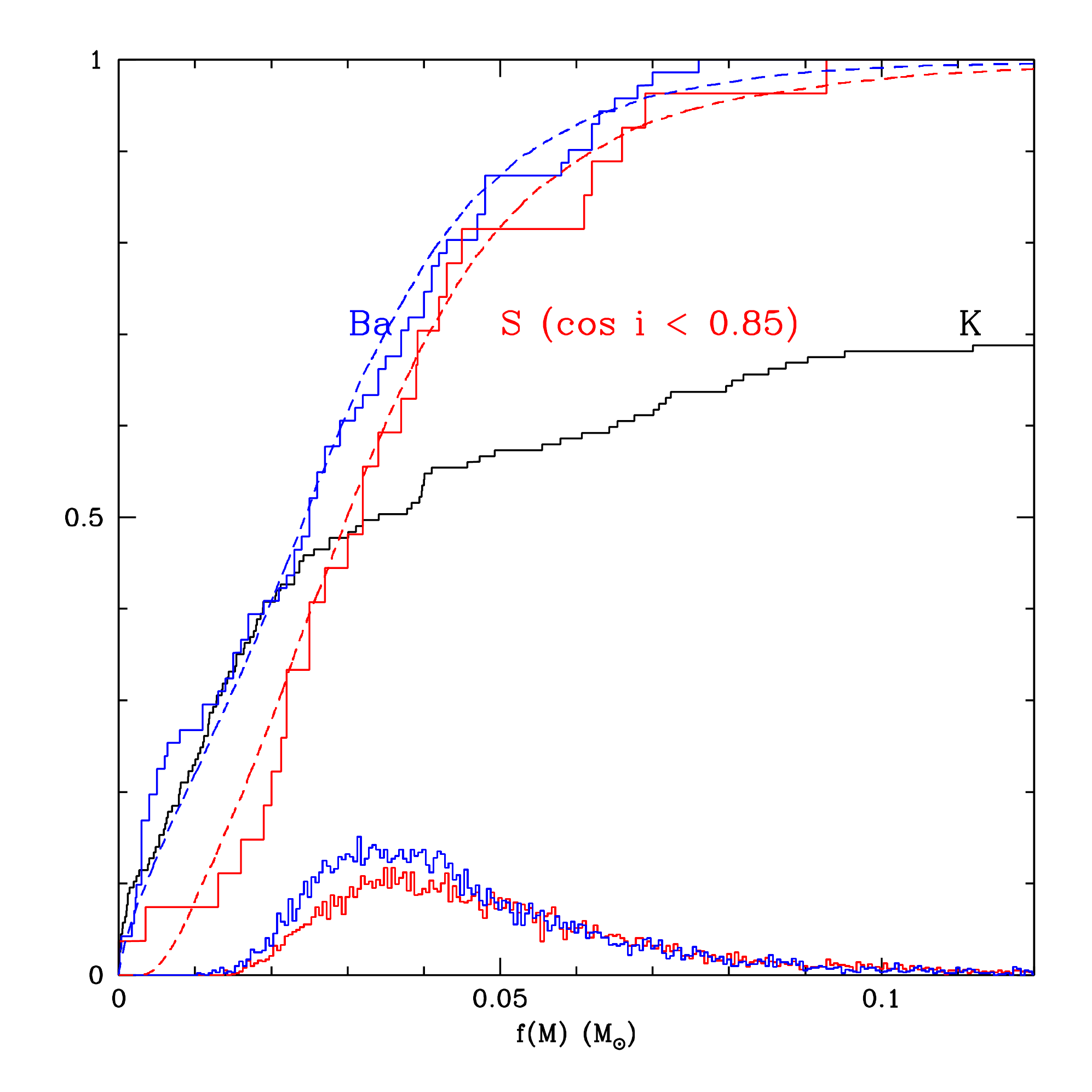}
\caption{\label{Fig:cumulative} Cumulative mass-function distributions for barium systems (blue), S-star systems (red), and K giants in clusters (black).  The step curves correspond to the observed values, as listed in Tables~\protect\ref{Tab:strongBa} -- \protect\ref{Tab:S}. The dashed curves are the synthetic curves produced by the adopted Gaussian distributions, with parameters as listed on the lines labelled Ba and S in Table~\protect\ref{Tab:Gau}.  The histograms at the bottom of the figure correspond to the histograms of $Q$  values for barium and S systems resulting from the normal distributions adopted for $M_1$ and $M_2$.
}
\end{figure}

In Fig.~\ref{Fig:bafq}, moreover, we see that there is a small excess of systems with mass ratios between 0.1 and 0.2; these systems would correspond to lower mass WDs, that is those below 0.46~M$_\odot$, namely, He WDs. Such WDs cannot have been the core of a thermally pulsing AGB star where the $s$-process material needed to pollute the barium star has been synthesised \citep{2016A&A...586A.151M}. Therefore,  this cannot be the correct explanation for the presence of that peak.  In fact, the peak may be traced to the presence of  the barium star HD~218356, =~56~Peg, with  a mass function of $(3.7\pm0.3) \times 10^{-5}$~M$_{\odot}$ , (Table~\ref{Tab:mildBa}) in the sample.  We checked that by removing 56~Peg from our sample, the small peak at $q \sim 0.15$ disappears.  A detailed analysis of that system  \citep{2006Obs...126...25F} concluded that to reconcile constraints from the orbital mass function with evolutionary considerations,   the giant must be a fast rotator ($V_{\rm rot}  \sim 30$ -- 50~km~s$^{-1}$) and the orbital inclination must be small ($i \sim 5^{\circ}$). This analysis led to masses in the range 2--4~M$_\odot$ and 0.75--1.15~M$_\odot$, for the primary and secondary respectively, corresponding to a mass ratio in the range 0.19--0.58, just above the values 0.1--0.2 inferred from the statistical analysis of the mass-function distribution. Since HD~218356 is classified as G8~Ib \citep{MKK}, the upper limit of the primary-mass range seems more likely, and yields  $q=0.19$. 
The small inclination of $5^{\circ}$ has a probability of occurrence of one part in 260, and is therefore not expected in a  sample of 71 stars such as ours. 
Therefore, we believe that the secondary peak observed at low $q$ values in Fig.~\ref{Fig:bafq} is the result of small-number  statistics and does not deserve any further discussion.

We performed the same analysis for our smaller sample of S star orbits ($N=29$) and the result is shown in Fig.~\ref{Fig:sfq}. We obtained a MRD similar to that of barium stars.
Both samples are combined in Fig.~\ref{Fig:basfq}, from which we conclude that S stars are the cooler analogues of barium stars since their orbital properties are similar, in agreement with the conclusion of  \citet{1998A&A...332..877J}.

\subsection{A different approach, fitting mass functions }
\label{Sect:sigma}

So far, we extracted the MRD from the inversion of the mass-function distribution using the Richardson-Lucy algorithm.  
Webbink (1988) and McClure \& Woodsworth (1990) have proposed a different approach, fitting the  mass-function distribution of barium stars by Gaussian distributions for $M_1$ and $M_2$ and a random distribution of orbital inclinations. 
After exploring the parameter space [$M_1$, $M_2$, $\sigma(M_1)$, $\sigma(M_2$)], the parameters minimizing the Kolmogorov-Smirnov distance between the observed and synthetic cumulative functions were considered as yielding the best fit, and were retained. Results from such a best fit of the $f(m)$ distribution are presented in Table~\ref{Tab:Gau} and Fig.~\ref{Fig:cumulative}, separately for barium and S systems.

In Table~\ref{Tab:Gau}, $M_{1}$ and $M_{2}$ denote the central values of the Gaussian distributions with standard deviations $\sigma_{1}$ and $\sigma_{2}$.  We stress that the fit  actually constrains $Q$, not $M_1$ and $M_2$ individually.  This is why Table~\ref{Tab:Gau} contains several ($M_1, M_2$) pairs, which all correspond to $Q = 0.041$~M$_{\odot}$ or $Q = 0.045$~M$_{\odot}$, the values yielding the best fit for barium and S stars, respectively. These values  agree well with the result previously obtained by \citet{1998A&A...332..877J} ($Q = 0.042\pm0.001$~M$_\odot$). 

S stars appear to be lacking small mass functions, and therefore to obtain a good  fit, one needs to assume  that the systems close to edge-on ($\cos i > 0.85$, or $i < 32^\circ$) are not present in the observed sample. This is not surprising, since small mass functions correspond to systems with small velocity amplitudes, which are difficult to detect if the systems  suffer from radial-velocity jitter, as it is the case for evolved giants like S stars. 

The smaller average mass for  S-star systems, as compared to barium systems, is expected since (extrinsic) S stars, which are  restricted to low surface temperatures, populate the tip of the RGB; temperature-wise, spectral type S is equivalent to M, and is
thus cooler than K. However, barium stars, with their earlier spectral types (K), could be a mix of RGB stars and He-clump stars because the latter are not  restricted to low-mass stars as are RGB stars.

A very interesting feature observed in Fig.~\ref{Fig:cumulative} is the distribution of the mass functions for the  binary K giants in open clusters from M07, which follows exactly that for barium stars,  before deviating for frequencies in excess of 43\%. This clearly means that at most 43\% of the M07 sample correspond to post-mass-transfer systems with WD companions. This frequency is an upper limit, since the low end of the $f(M)$ distribution merges WD and main-sequence companions, where the latter distribution becomes dominant for $f(M) > 0.02$~M$_\odot$. In any case, this result is consistent with the finding of Sect.~\ref{Sec:Merm} based on the inversion  of  the mass-ratio distribution that 22\% of the M07  binary sample corresponds to post-mass-transfer systems.
This prediction makes us suspect that some of the M07 stars should be rich in $s$ process elements as are the barium stars. This prediction is investigated in Sect.~\ref{Sect:abundances_all}.

\section{Are there barium stars among the Mermilliod binary giants in open clusters?}
\label{Sect:abundances_all}

In view of the difficulty in separating pre- from post-mass-transfer giants, from simple arguments based on the mass function and the location in the  ($P - e$) diagram (Sect.~\ref{Sect:post-mass-transfer}), we decided to perform an abundance analysis of those binaries falling in the same region of the ($P - e$) diagram as barium stars and with a mass function consistent with a WD companion. Such an analysis should uncover barium stars, if there are any hidden in the M07 sample. In the lower panel of Fig.~\ref{Fig:elogpmerm}, the promising candidates are  those   located below the green envelope. 
There are ten such stars that are observable from the Roque de los Muchachos Observatory (Canary Islands, Spain), where the HERMES spectrograph \citep{2011A&A...526A..69R} is mounted on the 1.2 m Mercator telescope from the Katholieke Universiteit Leuven. Two more were observed with the HARPS spectrograph on ESO 3.6 m telescope.
The target properties are listed in Table~\ref{Tab:targets}.

\begin{table*}
\caption[]{\label{Tab:targets}
Binary giants from the M07 sample that were the targets of an abundance analysis. As before, $Y \equiv f(M)/M_1$. The effective temperature $T_{\rm eff}$ and gravity $\log g$ are derived as explained in Sect.~\ref{Sect:parameters} or taken from the literature. The columns labelled TO and  Ba provide the cluster turn-off mass (in M$_\odot$) and  the conclusion of the abundance analysis, respectively: Y (barium star) or  N (not a barium star). In column 'spectro', HE stands for HERMES and HA for HARPS.
}
\begin{tabular}{lrllllrlllll}
\hline
Name & $P$ & $e$ & $\log Y$ & $T_{\rm eff}$ & $\log g$ & [Fe/H] & TO & Ba & Spectro & Ref\\ 
           & (d)  &       &               &   (K)             &                &            & (M$_\odot$)\\
\hline
\medskip\\
\noalign{Candidate post-mass-transfer binaries}
\medskip\\
\object{NGC 2539        209} & 11655  &0.16& -2.29 & 4750 & 2.5 & +0.13 &2.6 &N& HE &1\\
\object{NGC 2335        4}   & 301  &0.00& -2.46 & 4750 & 1.2 & -0.03    &4.3 & Y& HE &1\\
\object{IC 4756 139} & 3834        &0.22& -2.61  & 5220 & 2.7 & -0.06    &2.7 & Y& HE &1 \\
\object{NGC 2682        244} & 698         &0.11& -2.22  & 5150 & 2.6 & +0.00 &1.4 & N&HE &1 \\
\object{NGC 2682        170} & 4410        &0.50& -2.42  & 4250 & 1.7 & +0.00 &1.4 &N&HE &1 \\
\object{NGC 2682        143} & 43 &0.00& -2.75   & 5000 & 2.8 & +0.00 &1.4 &N&HE &1 \\
\object{NGC 2420        173} & 1479  &0.43& -2.42        & 5150 & 2.2 & -0.26 &2.1 &  Y & HE &1,2\\
\object{NGC 6940        111} & 3571        &0.30& -2.05  & 5150 & 2.9 & +0.01 &2.2 & N& HE &1\\
\object{NGC 2099        149} & 918         &0.13& -2.48  & 4900 & 2.1 & +0.08 &3.0 & N& HE &1 \\
\object{NGC 2099        966} & 3084        &0.37& -2.44  & 4600 & 2.1 & +0.08 & 3.0 &N?& HE &1 \\
\object{NGC 2477   1044} & 3108    &0.07&  -2.33  &  4780   &  2.6  & +0.01 &2.1 & Y & HA & 1\\
\object{NGC 4349     203} & 129.4   &0.028& -2.37  & 4940  &  2.2    & -0.07 & 3.2 & N & HA & 1\\
\medskip\\
\noalign{Binaries in the ($P - e$) gap}
\medskip\\
\object{IC 4756 69} & 1994 & $0.05\pm0.02$ & -2.07 & 4870 & 2.6 & -0.06 & 2.7 & N & HE & 1\\
\object{IC 4756 80} & 5791 & $0.03\pm0.01$ & -1.11 & 4910 & 2.8 & -0.06 & 2.7 & N & HE & 1\\
\object{NGC 2682 224} & 6645 & $0.01\pm0.01$&-1.04 & 4745 & 2.5 & 0.00&1.4  &N& HE &1\\
\medskip\\
\noalign{Search for barium stars in clusters, from the literature}
\medskip\\
\object{NGC 2420        250} & 1404 &0.08& -1.64 & -        & -     & -0.26 &2.1 & Y && 2\\
\object{NGC 5822        2}     & 1002  &0.13& -2.43& 5100  & 2.4  & -0.15 &2.1 &  Y && 3\\
\object{NGC 5822 201}    & ?           & ?   & ?      & 5200  & 2.7  & -0.15 &2.1 &  Y && 3\\
\object{NGC 5822 151}    & 1392   &0.21&-2.32& 4900     & 2.5     & -0.11 &2.1 &  N && 4\\
\medskip\\
\noalign{Reference star}
\medskip\\
\object{NGC 752 208} & 5214 & $0.13\pm0.01$ &-1.35 & 4760 & 2.5 & -0.08 & 1.7 & N &HE&1\\ 
\hline\\
\end{tabular}

References: (1) This work (2) \citet{2007A&A...470..919M} (3) \citet{2013AJ....146...39K}  (4) \citet{2014AJ....148...83S}
\end{table*}

On top of these 12 targets, 3 more were analysed because they fall in the long-period gap of the ($P - e$) diagram  (see Sect.~\ref{Sect:post-mass-transfer} for a description of this gap). The orbits were recomputed from the available radial-velocity data (as available on the WEBDA database), and with these recomputed eccentricities, one of  the binaries (IC 4756 69) is found to fall at the border of the ($P - e$) gap, but NGC 2682 224 and IC 4756 80 at least remain well within the gap (Table~\ref{Tab:targets}). On the other hand, only IC 4756 69 has a $\log Y$ value that makes it compatible with a post-mass-transfer object.
These three systems were added at the end of Table~~\ref{Tab:targets} as abundance targets as well. 

\begin{figure}
\includegraphics[width=9cm]{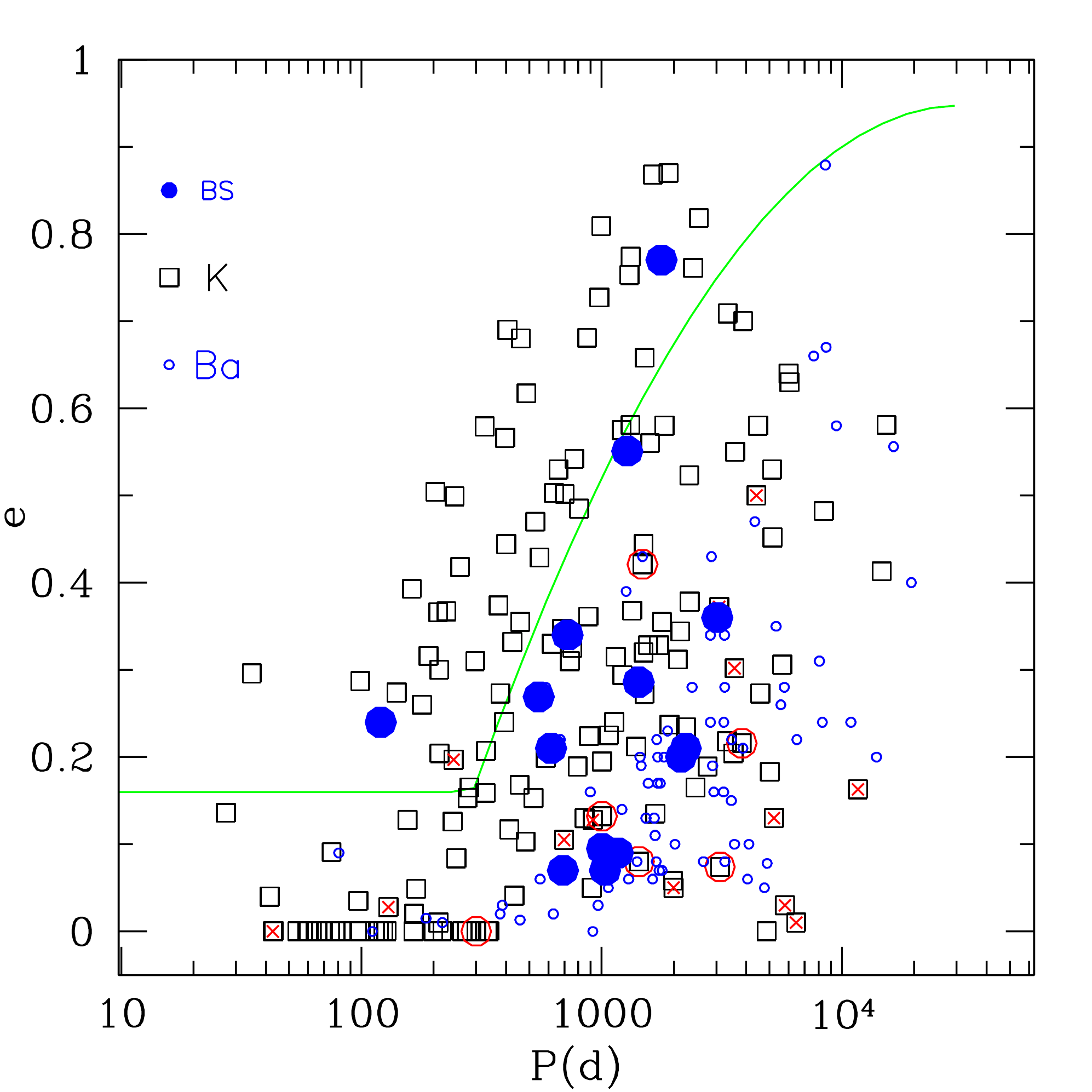}
\caption{\label{Fig:elogPBa}  
 Period - eccentricity diagram for the M07 sample of binary giants in open clusters (open black squares), barium stars (small open blue circles), and  blue-straggler binaries from the cluster NGC~188 \protect\citep[large filled blue dots;][]{2015ApJ...814..163G}.
M07 cluster giant stars  whose $s$-process abundances were derived are identified by red symbols: stars with normal $s$-process abundances are shown as crossed squares, and  stars with enhanced  $s$-process abundances are shown as circled squares, to  be compared with the location of barium stars.  The green solid line is the same envelope as that shown in Fig.~\protect\ref{Fig:elogpmerm} (see also Eq.~\protect\ref{Eq:ethresh}).
}
\end{figure}

Table~\ref{Tab:targets} also lists four barium stars in open clusters reported in the literature: NGC~2420 250, NGC 2420 173, NGC 5822        2, and NGC 5822 201.
The location in the $(P - e)$ diagram of all these cluster stars whose $s$-process abundances were derived are shown in Fig.~\ref{Fig:elogPBa}.

Finally, NGC 752 208 is observed to serve as a reference star, with no dynamical indication that it could be a post-mass-transfer object.

\subsection{Abundances}
\label{Sect:abundances}

For the sake of clarity, all details about the abundance analysis (e.g. derivation of stellar parameters and line lists) are provided in Appendix~\ref{Sect:abundances_appendix}. Here we provide only the final abundances (Table~\ref{Tab:final_abundances}).  The quoted errors on the abundances in that Table are only standard deviations around the mean from line-to-line scatter when several lines are available for a given chemical species. They do not include systematic errors from uncertainties on the model parameters (see Table~\ref{Tab:systematic_errors}  for the latter).

\begin{figure}
\begin{center}
\includegraphics[width=9cm]{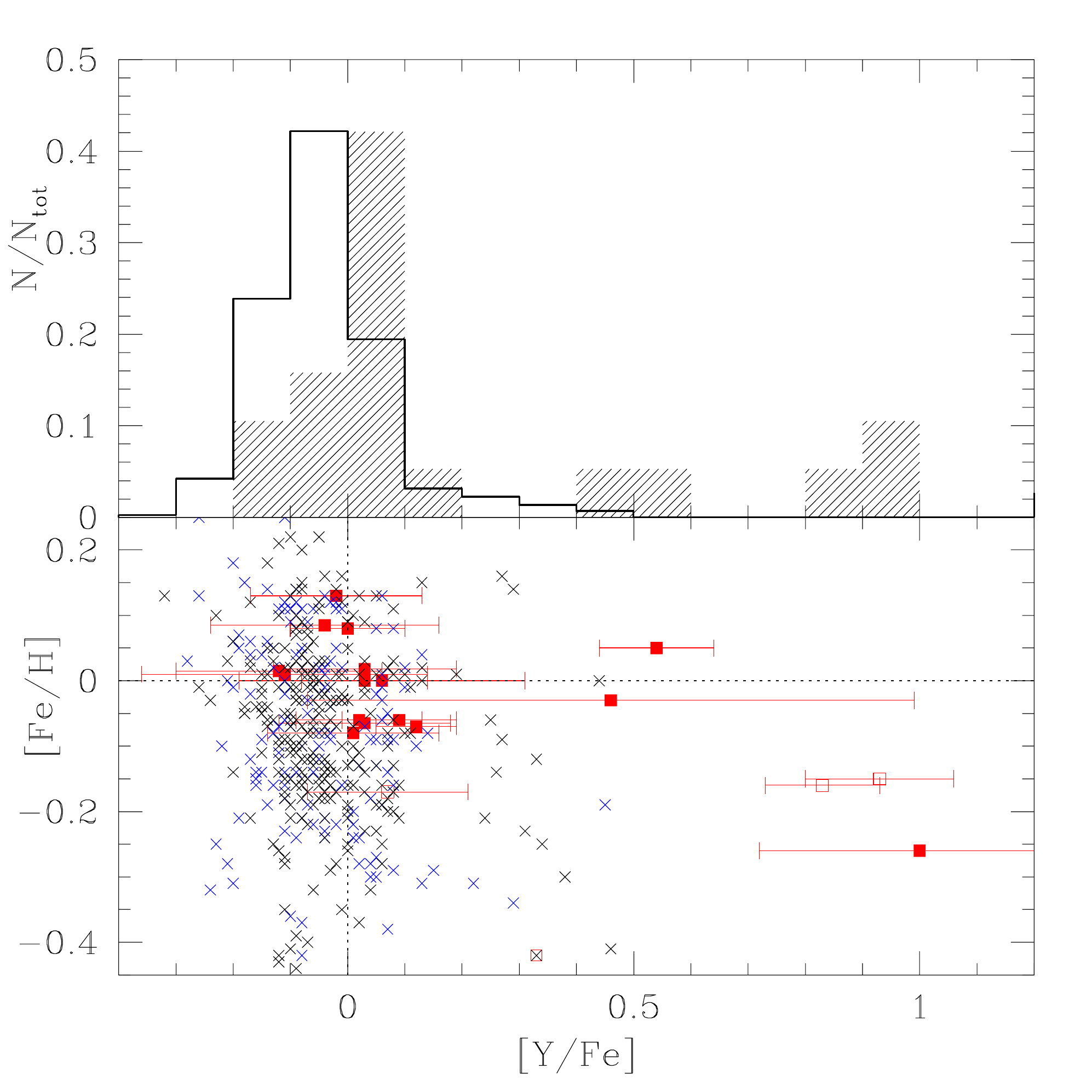}
\end{center}
\caption{\label{Fig:YFe}  Bottom panel: Relation between [Y/Fe] and metallicity [Fe/H] for cluster giants analysed in the present study (red filled squares). Literature data \citep[Stars 2, 151 and 201 in NGC~5822, from][]{2013AJ....146...39K,2014AJ....148...83S} are depicted by red open squares. For the sake of clarity, stars with identical metallicities have been slightly shifted horizontally. Field giants, from \citet{Luck_Heiter_2007} and \citet{2007ARep...51..382M}, are depicted by small (respectively black and blue) crosses. The barium star HD~104979 is part of  the \citet{Luck_Heiter_2007} sample and is identified by a crossed square. Top panel: The normalised  distribution of [Y/Fe] for the field giants is represented by the unshaded histogram. The shaded histogram corresponds to the cluster giants (including some barium stars) from the present study.
}
\end{figure}

\begin{figure}
\begin{center}
\includegraphics[width=9cm]{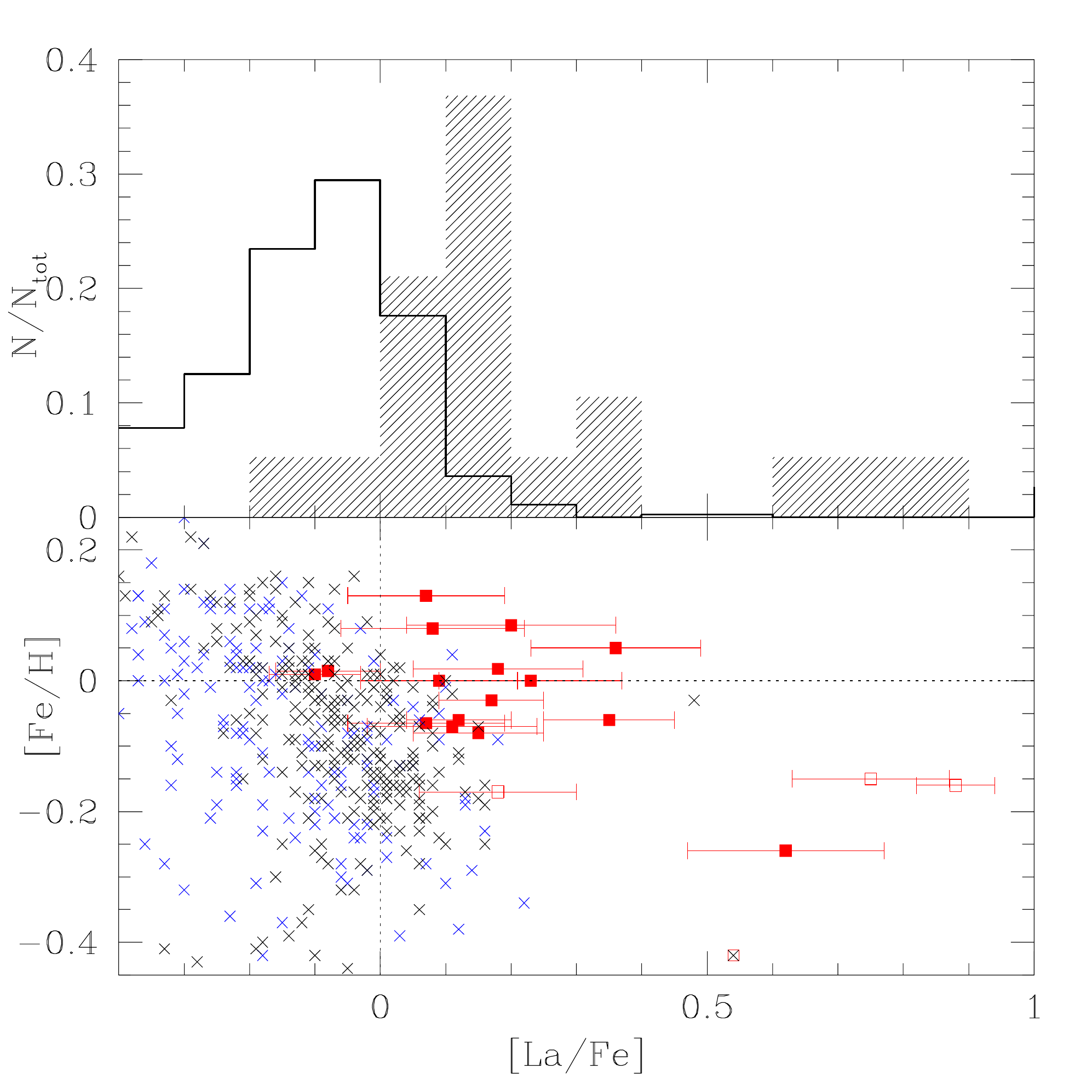}
\end{center}
\caption{\label{Fig:LaFe} 
Same as Fig.~\ref{Fig:YFe} for La.
}
\end{figure}

\begin{figure}
\begin{center}
\includegraphics[width=9cm]{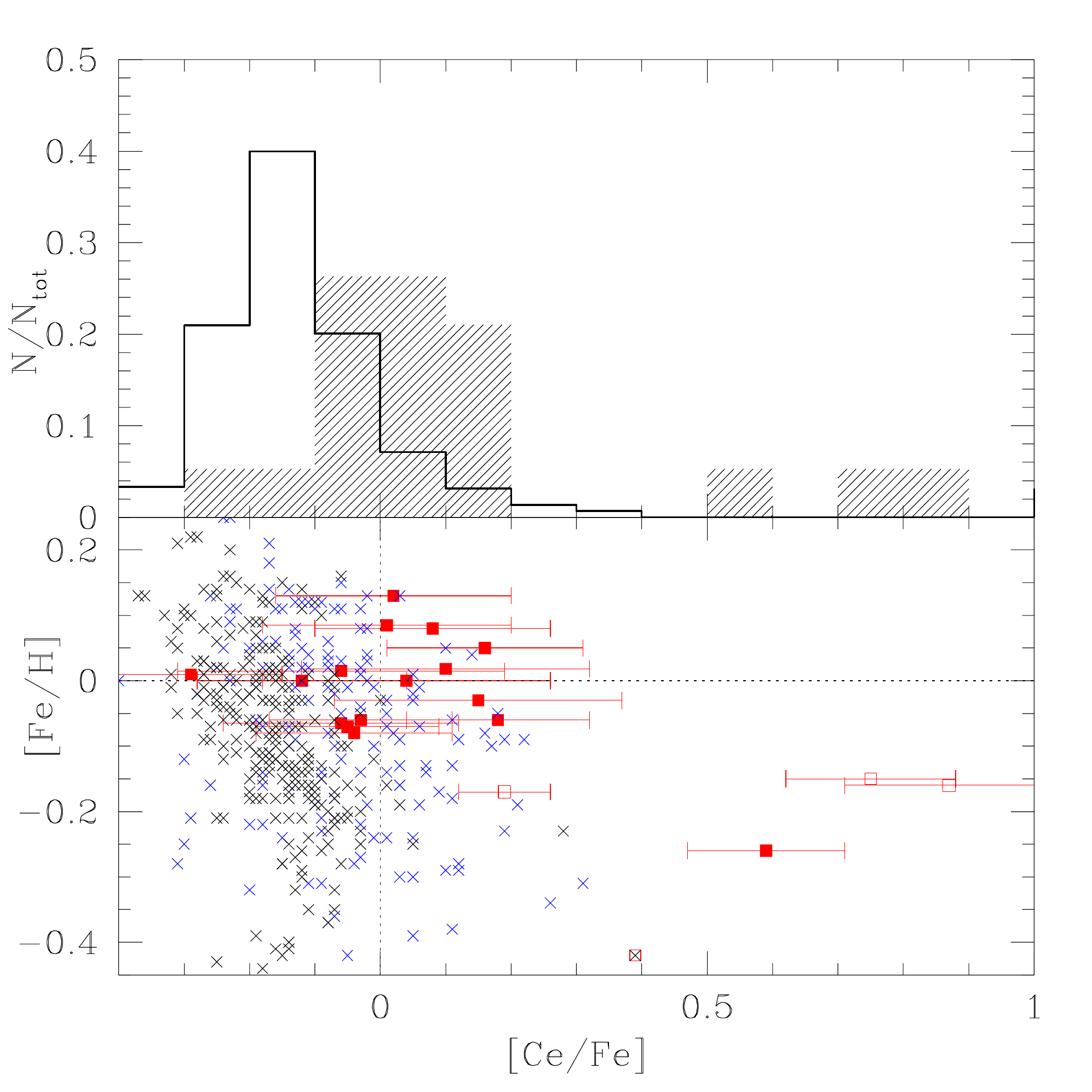}
\end{center}
\caption{\label{Fig:CeFe} 
Same as Fig.~\ref{Fig:YFe} for Ce.
}
\end{figure}

\begin{figure}
\begin{center}
\includegraphics[width=9cm]{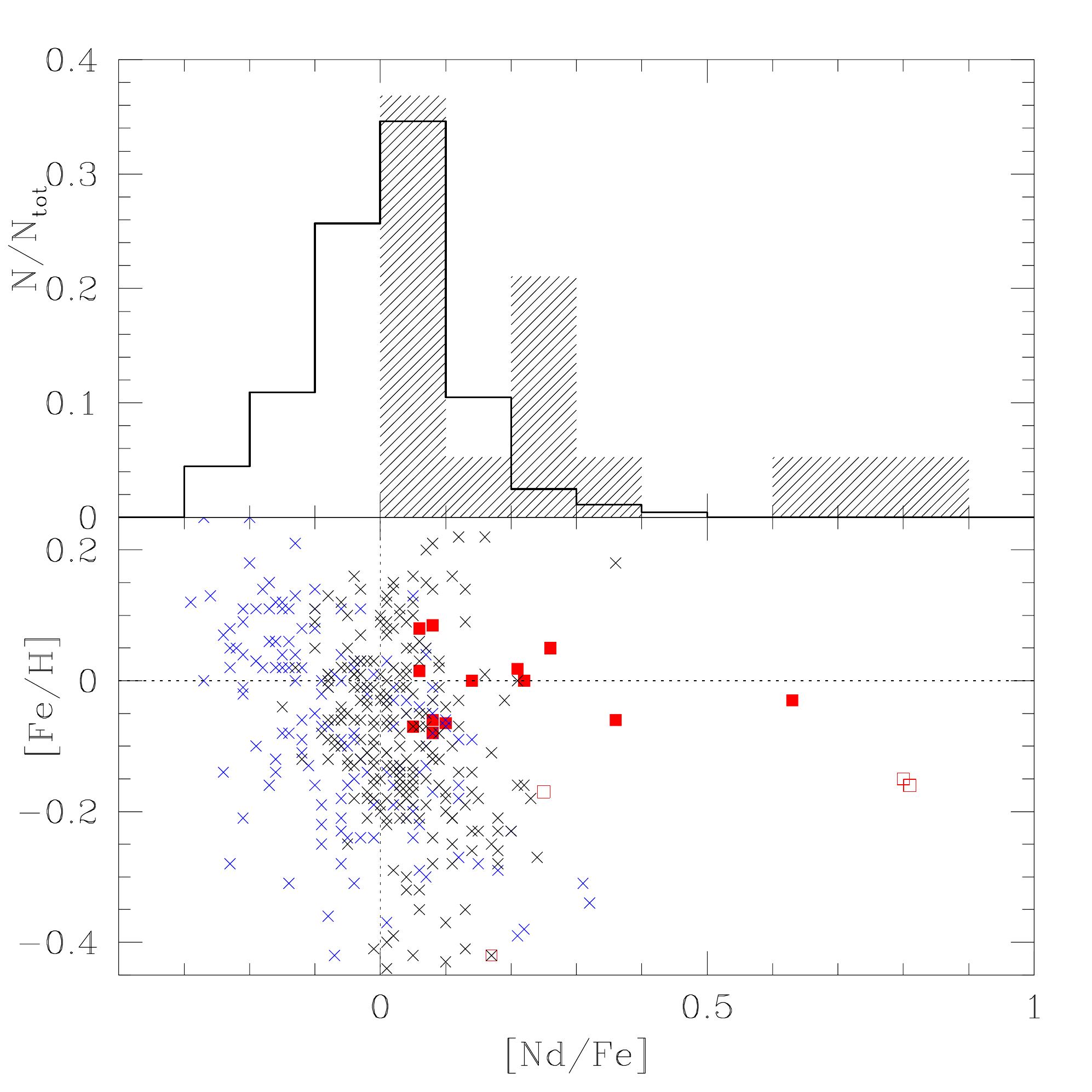}
\end{center}
\caption{\label{Fig:NdFe} 
Same as Fig.~\ref{Fig:YFe} for Nd. There is no error bar available on the Nd abundance, which is derived from single line.
}
\end{figure}

The abundances in Table~\ref{Tab:final_abundances}
reveal that none of the three stars analysed in the long-period gap in the ($P - e$) diagram are  barium stars nor is the comparison star NGC~752~208. Among the 12 targets with post-mass-transfer properties:  one is a strong barium star (NGC~2420 173; see Sect.~\ref{Sec:Ba} for the definition of mild and strong barium stars), which is a fact already noted by \citet{2007A&A...470..919M}, but without any quantitative analysis; and  three show mild overabundances for at least two $s$-process elements among the five studied (Y, Zr, La, Ce, Nd). The threshold for flagging a star as a mild barium star may be best evaluated  using as reference the abundance dispersion of the sample of field red giants studied by  \citet{Luck_Heiter_2007} and \citet{2007ARep...51..382M}, which are shown as small crosses in Figs.~\ref{Fig:YFe}  -- \ref{Fig:NdFe} (and non-shaded histogram).   
These four figures compare the distributions of the [Y/Fe], [La/Fe], [Ce/Fe], and [Nd/Fe]  abundance ratios   in field giants and in the cluster giants targeted for the abundance analysis. From the abundance distribution  in field stars, we may assess that [X/Fe]~$ > 0.25$~dex (where X stands for any of  Y, La, Ce, or Nd) represents a reasonable threshold to flag a star as a barium star, since very few field giants venture into that region.   
Based on this criterion, the following three stars are outliers with respect to the field giants for at least two elements  (mentioned between parentheses in the following list), and may thus be considered as mild barium stars: IC~4756 139 (Zr, La, Ce, Nd), NGC~2335 4 (Y, Nd), and NGC~2477 1044 (Y, Zr, La, Nd), and eight are not enriched ([X/Fe]~$< 0.25$~dex) in $s$-process elements (NGC~2099 149, NGC~2099 966, NGC~2539 209, NGC~2682 143, NGC~2682 170, NGC~2682 244, NGC~4349 203,  and NGC~6940 111). Thus, among the cluster giants with a mass function that is compatible with a WD companion, 
33\% (=4/12) show a chemical signature of mass transfer in the form of  $s$-process overabundances. None of the three stars in the long-period gap turn out to be a barium star.
 
\begin{figure}
\begin{center}
\includegraphics[width=9cm]{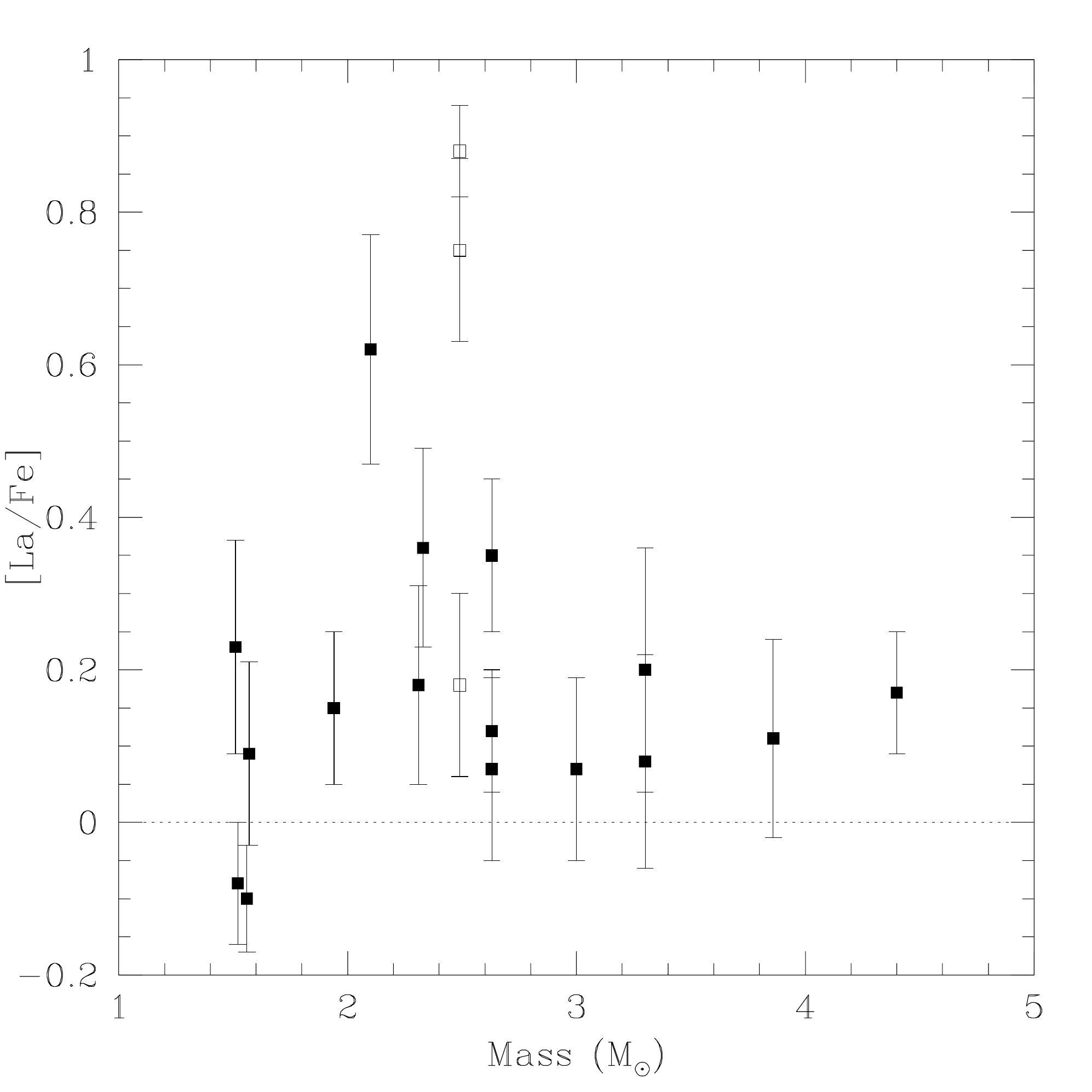}
\end{center}
\caption[]{\label{Fig:La-TO}
Relation between [La/Fe] and the mass of the giant, taken from Table~\protect\ref{Tab:cluster_mass}.  Open symbols denote data taken from literature, from the reference listed in Table~\protect\ref{Tab:targets}, and filled symbols denote abundances from the present  analysis.
}
\end{figure}

Moreover, as shown by Figs.~\ref{Fig:YFe} -- \ref{Fig:NdFe}, the degree of overabundance seems correlated with the cluster metallicity: the star NGC 2420 173, which belongs to the most metal-poor cluster of the sample ([Fe/H]~=~$-0.26$~dex), exhibits strong $s$-process overabundances, for elements from the first and second $s$-process peaks. Incidentally, NGC~2420 contains another barium star, NGC~2420 250 \citep{2007A&A...470..919M}. Two among the mild barium stars are found in the 
clusters with a slightly subsolar metallicity,  IC~4756 and NGC~2335 (with $-0.10 \le$~[Fe/H]~< 0.0), as is the case for the barium stars found in NGC~5822 ([Fe/H] = -0.15~dex) by \citet{2013AJ....146...39K}  and  \citet{2014AJ....148...83S}, whereas the fraction of mild barium stars is very low among solar-metallicity clusters (only one is found in NGC~2477, with [Fe/H] = 0.01~dex).   
It is tempting to attribute this correlation to    the sensitivity of the $s$-process efficiency with metallicity, as predicted, for example, by  \citet{2000A&A...362..599G}. Another explanation, in terms of the dilution factor  of the accreted matter in the giant's envelope, is not  supported by our results (Fig.~\ref{Fig:La-TO}). This is because   in the cluster with the lowest turn-off (TO) mass, NGC 2682, none of the three stars studied turn out to be a barium star (even though NGC 2682 143, with [La/Fe]~=~0.23~dex, is a close call).
Therefore, the dilution factor is not the main factor controlling the level of  $s$-process abundances in this cluster where giants have  a small envelope mass and, thus, it should be easy to form barium stars.  Overall, there is no correlation between [La/Fe] and the TO mass, as shown in Fig.~\ref{Fig:La-TO}.

A last comment regarding the TO mass concerns the cluster NGC~2335, which has a large TO mass of 4.3~M$_\odot$. It is remarkable that it hosts a mild barium star, as diagnosed from its [Y/Fe] and [Nd/Fe] abundances (+0.46 and +0.63~dex, respectively); this means that rather massive AGB stars are still able to operate the $s$ process.

\begin{table*}
\setlength{\tabcolsep}{3pt} 
  \begin{center}
    \caption{\label{Tab:final_abundances} Final $\abratio{X}{Fe}$ abundance ratios for Y, Zr, La, Ce, and Nd for open cluster stars, normalised by the stellar metallicity.  The quoted errors on the abundances are only standard deviations around the mean from line-to-line scatter when several lines are available for a given chemical species. They do not include systematic errors from uncertainties on the model parameters (see Table~\ref{Tab:systematic_errors}  for the latter).
The integer number following the abundances corresponds the number of lines used to compute the mean.
The last column flags the star according to the following label: N (not a barium star), M (mild barium star), and S (strong barium star).
}
    
\begin{tabular}{lrlrlllllrllllll}
\hline\\ 
             &  [Fe/H]& \multicolumn{2}{c}{[Y II/Fe]} & \multicolumn{2}{c}{[Zr I/Fe]} & \multicolumn{2}{c}{[Zr II/Fe]} & \multicolumn{2}{c}{[LaII/Fe]} & \multicolumn{2}{c}{[CeII/Fe]} & \multicolumn{2}{c}{[NdII/Fe]} & Ba\\
\hline
\medskip\\
Sun          & 0.0   &$-0.01\pm0.02$&10&$ +0.03\pm0.06$&3&$-0.01$&1&$ +0.04\pm0.07$& 9&$-0.03\pm0.10$&5&$-    $&0&-\\
IC 4756 139   & -0.06 &$+0.02\pm0.11$& 9&$ +0.34\pm0.07$&3&$-    $&0&$ +0.35\pm0.10$&12&$+0.18\pm0.14$&8&$+0.36$&1&M\\
NGC 2099 149  & +0.08 &$+0.00\pm0.10$& 9&$ -0.03\pm0.04$&2&$-0.18$&1&$ +0.08\pm0.14$&10&$+0.08\pm0.18$&7&$+0.06$&1&N\\
NGC 2099 966  & +0.08 &$-0.02\pm0.20$& 9&$ -0.20\pm0.03$&3&$-$    &0&$ +0.20\pm0.16$&12&$+0.01\pm0.19$&6&$+0.08$&1&N\\
NGC 2335 4    & -0.03 &$+0.46\pm0.53$&10&$ -$           &0&$-$    &0&$ +0.17\pm0.08$& 6&$+0.15\pm0.22$&5&$+0.63$&1&M\\
NGC 2420 173  & -0.26 &$+1.00\pm0.28$&10&$ +0.72\pm0.00$&3&$-$    &0&$ +0.62\pm0.15$& 7&$+0.59\pm0.12$&3&$+1.51$&1&S\\
NGC 2539 209  & +0.13 &$-0.02\pm0.15$& 9&$ -0.32\pm0.02$&2&$-0.10$&1&$ +0.07\pm0.12$& 9&$+0.02\pm0.18$&6&$-    $&0&N\\
NGC 2682 143  & +0.00 &$+0.06\pm0.25$& 6&$ +0.17\pm0.09$&3&$-$    &0&$ +0.23\pm0.14$& 4&$+0.04\pm0.22$&5&$+0.22$&1&N\\
NGC 2682 170  & +0.00 &$-0.11\pm0.25$& 6&$ -0.26\pm0.11$&4&$-0.23$&1&$ -0.10\pm0.07$&10&$-0.29\pm0.14$&4&$-    $&0&N\\
NGC 2682 244  & +0.00 &$-0.12\pm0.18$&11&$ +0.17\pm0.09$&4&$-0.28$&1&$ -0.08\pm0.08$&10&$-0.06\pm0.25$&6&$+0.06$&1&N\\
NGC 6940 111  & +0.01 &$+0.03\pm0.16$&10&$ +0.39\pm0.07$&3&$-0.10$&1&$ +0.18\pm0.13$&13&$+0.10\pm0.22$&7&$+0.21$&1&N\\
IC 4756  69  & -0.06 &$+0.09\pm0.10$& 7&$ -0.02\pm0.47$&4&$+0.27\pm0.11$&3&$+0.12\pm0.08$&12  &$-0.03\pm0.14$&9&$+0.08$&1&N \\
IC 4756  80  & -0.06 &$+0.03\pm0.15$& 6&$ -0.02\pm0.48$&4&$ +0.14\pm0.09$&3&$ +0.07\pm0.12$&11&$ -0.06\pm0.18$& 9&$+0.10$&1&N\\
NGC 2477 1044 & +0.01 &$+0.54\pm0.10$& 9&$ +0.29\pm0.31$&9&$ +0.65\pm0.14$&3&$ +0.36\pm0.13$&14&$ +0.16\pm0.15$& 9&$+0.26$&1&M\\
NGC 2682 224  &  0.00 &$+0.03\pm0.11$& 7&$ +0.20\pm0.36$&8&$ +0.06\pm0.15$&3&$ +0.09\pm0.12$&13&$ -0.12\pm0.16$&10&$+0.14$&1&N\\
NGC 4349 203  & -0.07 &$+0.12\pm0.07$& 9&$ +0.03\pm0.48$&5&$ +0.26\pm0.16$&3&$ +0.11\pm0.13$&14&$ -0.05\pm0.14$& 8&$+0.05$&1&N\\
NGC 752 208  & -0.08 &$+0.01\pm0.15$& 8&$ +0.22\pm0.40$&8&$ +0.14\pm0.12$&3&$ +0.15\pm0.10$&15&$ -0.04\pm0.15$&10&$+0.08$&1&N\\
\hline
\end{tabular}
  \end{center}
\end{table*}

\subsection{Discussion}

The discovery of barium or S stars in open clusters is a real golden nugget as they represent  a very definite signature of post-mass-transfer objects. \citet{2013AJ....146...39K} report the existence of two barium stars in the open cluster NGC 5822, one of which is among the sample of M07: NGC 5822~2 is known to have an orbital period of 1002.2 days and an eccentricity of 0.132. As such, it does follow the trend shown by barium and S stars to have smaller eccentricities than normal giants at a given orbital period. 
        However, in the same cluster, the M07 sample contains two other spectroscopic binaries:  NGC 5822~151 and NGC 5822~276.  \citet{2014AJ....148...83S} showed that the former is not a barium star. This binary  has a period of 1391.5 d and an eccentricity of 0.212.
% and the second one has $P=6048$~d and $e=0.63$ and
It does therefore fall within the (field) barium star locus  in the $(P - e)$ diagram, despite not being a barium star, thus confirming the conclusion already apparent from Fig.~\ref{Fig:elogPBa} that M07 cluster giants contain a mix of barium and non-barium stars at long periods and small eccentricities.

        The barium star NGC 5822~2 has a spectroscopic mass function of $f(m)=0.00916$~M$_\odot$ and an estimated mass of 2.49~M$_\odot$, i.e. $\log Y=-2.43$, putting it right in the middle of the systems containing a WD. This is thus no surprise. However, this is also the case of NGC 5822~151 ($\log Y=-2.15$), which is not a barium star.

Our abundance study described in Sect.~\ref{Sect:abundances}, involving 12 more giants from the M07 sample, has generalised the above conclusion that the locus of barium stars in the $(P-e)$ diagram also contains non-barium stars. It is unfortunately not possible to infer the nature of the companion (WD or low-mass main-sequence star)  so as to decide whether the non-barium nature of these stars is a consequence of their pre-mass-transfer status or a consequence of mass transfer of matter that is not enriched in $s$ process.
 The statistics of barium stars found in clusters may nevertheless be used to shed light on that question.
The 12 giants from M07 subjected to an abundance study cover the range $-2.75 \le \log Y \le -2.05$ (Table~\ref{Tab:targets}). In that range, our statistical analysis of the reduced mass functions $Y$ predicts 21 systems with WD companions and 16 with main-sequence companions (see Fig.~\ref{Fig:mermfmall}), or a percentage of 57\% (= 21/37) for the former\footnote{This percentage is larger than the overall 22\% fraction discussed earlier because of the pre-selected range in $\log Y$.}. Applied to the sample of 12 red giants whose $s$-process abundances were derived, this percentage would imply that $0.57 \times 12 = 6.8\pm1.4$ barium stars should have been found if all systems with WD companions were barium stars;  the uncertainty on that value is estimated from the hypergeometric distribution, with $N = 37$, $N_1 = 12$, and $p = X/N = 21/37 = 0.57$. Only four were found, suggesting (despite the small-number statistics) that not all mass-transfer events from an AGB companion lead to the formation of a  barium star.

Another signature of mass transfer is the blue straggler phenomenon. \citet{2011Natur.478..356G}, \citet{2014ApJ...783L...8G}, and \citet{2015ApJ...814..163G} have  shown that (most of the) blue straggler stars in the cluster NGC~188 are binary systems, some with detected WD companions. The distribution of these binaries in the ($P - e$) diagram is shown in Fig.~\ref{Fig:elogPBa}, using the orbital elements from  \citet{2015ApJ...814..163G}. Surprisingly, considering the fact the blue stragglers are supposed to be  post-mass-transfer objects, some of these objects are located outside the region defined by the post-mass-transfer barium stars.  Unfortunately, a possible barium overabundance in the blue-straggler binaries of NGC~188 has not been tested, as was carried out by \citet{2015AJ....150...84M} for the blue stragglers in NGC~6819. In the latter cluster, five blue stragglers are found to be enriched in barium. Among these however, four show no sign of  binarity and one is a SB2 system; neither situation is compatible with the mass-transfer scenario of matter rich in $s$-process elements from an AGB companion to produce such barium-rich stars. Overall, the relation between barium stars and blue stragglers has thus not yet been convincingly established.

\section{Conclusions}

This paper presents a study of the mass-function distributions of red giants in open clusters and in barium systems. As far as the cluster sample is concerned, we conclude that 22\% of the sample corresponds to post-mass-transfer systems. These systems must have WD companions. An abundance study of  12 cluster giant stars with a reduced mass function compatible with such a WD companion 
reveals that only 4 are indeed barium stars (3 are mild barium stars, and 1 is a strong barium star), whereas the statistics predicts 7 barium stars if all mass-transfer events ending up with a WD companion would also contaminate the accreting star with $s$-process elements and turn it into a barium star.  We are thus led to conclude that this is not always the case. This lack of $s$-process contamination is probably related to the high metallicity of the considered clusters (most have solar, or even slightly super-solar metallicity), since all strong barium stars are found in the clusters with the lowest metallicities. This result confirms the larger efficiency of the $s$-process nucleosynthesis predicted at lower metallicities.

An important result of our analysis is the fact that some post-mass-transfer systems from the M07 sample of giants in open clusters  are located outside the locus of barium systems in the $(P-e)$ diagram. The same conclusion holds for the blue-straggler binaries in the cluster NGC 188. The origin of this difference in the  dynamical outcome of mass transfer in barium stars, on one hand, and blue-straggler or open-cluster systems, on the other hand, is so far unknown. 

Regarding barium stars, we confirm earlier results that their mass-function distribution is typical of  WD companions, but in the absence of a reliable mass distribution for the barium giants,  it is not possible to compare the WD mass distribution in barium systems with that of field WDs.

\begin{acknowledgements}
MvdS and SvE are supported by a grant from the {\it Fondation ULB}. Based on observations obtained with the HERMES spectrograph, which is supported by the Research Foundation - Flanders (FWO), Belgium, the Research Council of KU Leuven, Belgium, the Fonds National de la Recherche Scientifique (F.R.S.-FNRS), Belgium, the Royal Observatory of Belgium, the Observatoire de Gen\`eve, Switzerland and the Th\"{u}ringer Landessternwarte Tautenburg, Germany. We thank the anonymous referee for useful comments.
\end{acknowledgements}

\appendix
\section{Mass functions, periods, and eccentricities for barium and S stars}

Since the review by \citet{1998A&A...332..877J}, several more orbits for barium and S stars have appeared in the literature. We therefore considered it useful to collect all orbits available in Tables~\ref{Tab:strongBa} (strong barium stars), \ref{Tab:mildBa} (mild barium stars), and \ref{Tab:S} (S stars).

\begin{table*}[ht]
\caption[]{
\label{Tab:strongBa}
Table of mass functions for barium stars 
with strong chemical anomalies used in the current study, along with orbital periods and eccentricities. 
The systems are ordered by increasing orbital period.
References are provided at the end of Table~\ref{Tab:S}. If not specified otherwise, the column 'Name' lists the HD number.
}
\begin{tabular}{llllllll}
\multicolumn{1}{c}{Name} & \multicolumn{1}{c}{Period (d)} & \multicolumn{1}{c}{$e$} & 
\multicolumn{1}{c}{$f(M)$ (M$_{\odot}$)} & Ref.\\
\hline\\
\object{121447} & $185.7\pm0.1$ & $0.01\pm0.01$ & $0.025\pm0.001$ & 1\tabularnewline
\object{120620} & $217.2\pm0.1$ & $0.01\pm0.01$ & $0.062\pm0.001$ & 4\tabularnewline
\object{+38$^\circ$118}(1+2) & $299.4\pm0.2$ & $0.14\pm0.01$ & $0.014\pm0.0004$ & 4\tabularnewline
\object{24035} & $377.8\pm0.3$ & $0.02\pm0.01$ & $0.047\pm0.003$ & 4\tabularnewline
\object{-64$^\circ$4333} & $386.0\pm0.5$ & $0.03\pm0.01$ & $0.068\pm0.003$ & 4\tabularnewline
\object{46407} & $457.4\pm0.1$ & $0.013\pm0.008$ & $0.035\pm0.001$ & 5\tabularnewline
\object{100503} & $554.4\pm1.9$ & $0.06\pm0.05$ & $0.011\pm0.001$ & 4\tabularnewline
\object{199939} & $584.9\pm0.7$ & $0.28\pm0.01$ & $0.025\pm0.001$ & 6\tabularnewline
\object{44896} & $628.9\pm0.9$ & $0.02\pm0.01$ & $0.048\pm0.0015$ & 5\tabularnewline
\object{92626} & $918.2\pm1.2$ & $0.00\pm0.01$ & $0.042\pm0.002$ & 5\tabularnewline
\object{Lu163} & $965\pm15$ & $0.03\pm0.07$ & $0.0029\pm0.0006$ & 4\tabularnewline
\object{211594} & $1018.9\pm2.7$ & $0.06\pm0.01$ & $0.0140\pm0.0005$ & 5\tabularnewline
\object{31487} & $1066.4\pm2.6$ & $0.05\pm0.01$ & $0.038\pm0.002$ &  6\tabularnewline
\object{NGC2420-250} (X) & $1403.6\pm3.5$ & $0.08\pm0.03$ & $0.047\pm0.005$ & 7\tabularnewline
\object{88562} & $1445.0\pm8.5$ & $0.20\pm0.02$ & $0.048\pm0.003$ & 4\tabularnewline
\object{NGC2420-173} & $1479\pm9$ & $0.43\pm0.05$ & $0.008\pm0.002$ & 7\tabularnewline
\object{84678} & $1630\pm10$ & $0.06\pm0.02$ & $0.062\pm0.003$ & 4\tabularnewline
\object{154430} & $1668\pm17$ & $0.11\pm0.03$ & $0.034\pm0.003$ & 4\tabularnewline
\object{43389} & $1689\pm9$ & $0.08\pm0.02$ & $0.043\pm0.002$ & 5\tabularnewline
\object{101013} & $1711\pm4$ & $0.20\pm0.01$ & $0.037\pm0.001$ & 6,8 \tabularnewline
\object{201657} & $1710\pm15$ & $0.17\pm0.07$ & $0.004\pm0.001$ & 5\tabularnewline
\object{49641} & $1768\pm23$ & $0.07\pm0.11$ & $0.0031\pm0.0004$ &6,21 \tabularnewline
\object{5424} & $1882\pm19$ & $0.23\pm0.04$ & $0.005\pm0.0004$ & 4 \tabularnewline
\object{16458} & $2018\pm12$ & $0.10\pm0.02$ & $0.041\pm0.003$ & 1\tabularnewline
\object{20394} & $2226\pm22$ & $0.20\pm0.03$ & $0.0020\pm0.0005$ & 9\tabularnewline
\object{36598} & $2653\pm23$ & $0.08\pm0.02$ & $0.037\pm0.002$ & 4\tabularnewline
\object{178717} & $2866\pm21$ & $0.43\pm0.03$ & $0.006\pm0.001$& 6\tabularnewline
\object{201824} & $2837\pm13$ & $0.34\pm0.02$ & $0.040\pm0.003$& 9\tabularnewline
\object{50082} & $2896\pm21$ & $0.19\pm0.02$ & $0.027\pm0.002$ & 5\tabularnewline
\object{42537} & $3216\pm55$ & $0.16\pm0.05$ & $0.027\pm0.005$ & 4\tabularnewline
\object{-42$^\circ$2048} & $3260\pm28$ & $0.08\pm0.02$ & $0.065\pm0.004$ & 4\tabularnewline
\object{107541} & $3570\pm46$ & $0.10\pm0.03$ & $0.029\pm0.002$ & 5\tabularnewline
\object{+38$^\circ$118}(12+3) & $3877\pm112$ & $0.21\pm0.06$ & $0.0017\pm0.0004$ & 4\tabularnewline
\object{196445} & $3221\pm43$ & $0.24\pm0.02$ & $0.031\pm0.002$ &4\tabularnewline
\object{60197} & $3244\pm66$ & $0.34\pm0.05$ & $0.0028\pm0.0006$ & 4\tabularnewline
\object{123949} & $8539\pm25$ & $0.88\pm0.01$ & $0.048\pm0.007$ & 0\tabularnewline
\object{211954} & $10908\pm164$ & $0.24\pm0.06$ & $0.076\pm0.071$  & 0\tabularnewline
\object{65854} & SB &  & & 0\tabularnewline
\object{19014} & cst? & & & 10\tabularnewline
\hline
\medskip\\
\end{tabular}
\end{table*}

\begin{table*}[ht]
\caption{\label{Tab:mildBa}
Same as Table~\ref{Tab:strongBa} for Ba stars with mild chemical anomalies.}
\begin{tabular}{lllllllll}
\multicolumn{1}{c}{Name} & \multicolumn{1}{c}{Period (d)} & \multicolumn{1}{c}{$e$} & 
\multicolumn{1}{c}{$f(M)$ (M$_{\odot}$)} & Ref.\\
\hline\\
\object{77247} & $80.53\pm0.01$ & $0.09\pm0.01$ & $0.0050\pm0.0001$ & 6\tabularnewline
\object{218356} & $111.16\pm0.02$ & 0. & $0.000037\pm0.000003$ & 11\tabularnewline
\object{136138} & $506.4\pm0.2$ & $0.333\pm0.006$ &$0.0113\pm0.0003$ & 3,14 (HR 5692) \\
\object{58368} & $672.7\pm1.3$ & $0.22\pm0.02$ & $0.021\pm0.001$ & 6\tabularnewline
\object{49841} & $896\pm2$ & $0.15\pm0.02$ & $0.032\pm0.002$ & 5\tabularnewline
\object{58121} & $1214.3\pm5.7$ & $0.14\pm0.02$ & $0.015\pm0.001$ & 5\tabularnewline
\object{26886} & $1263.2\pm3.7$ & $0.39\pm0.02$ & $0.025\pm0.002$ & 5\tabularnewline
\object{223617} & $1293.7\pm3.9$ & $0.06\pm0.02$ & $0.0064\pm0.0004$ &5,6\tabularnewline
\object{143899} & $1461.6\pm6.9$ & $0.19\pm0.02$ & $0.017\pm0.001$ & 4\tabularnewline
\object{210946} & $1529.5\pm4.1$ & $0.13\pm0.01$ & $0.041\pm0.001$ & 5\tabularnewline
\object{101079} & $1563\pm2$ & $0.171\pm0.005$ & $0.00232\pm0.00005$ & 0\\
\object{95193} & $1653.7\pm9.0$ & $0.13\pm0.02$ & $0.026\pm0.001$ & 4\tabularnewline
\object{27271} & $1693.8\pm9.1$ & $0.22\pm0.02$ & $0.024\pm0.001$ &5\tabularnewline
\object{200063} & $1735.4\pm8.1$ & $0.07\pm0.04$ & $0.058\pm0.004$ &5\tabularnewline
\object{91208} & $1754\pm13$ & $0.17\pm0.02$ & $0.022\pm0.002$ & 4\tabularnewline
\object{288174} & $1817.5\pm6.7$ & $0.20\pm0.01$ & $0.017\pm0.001$ & 4\tabularnewline
\object{204075} & $2378\pm55$ & $0.28\pm0.07$ & $0.004\pm0.001$ & 6,10\tabularnewline
\object{205011} & $2837\pm10$ & $0.24\pm0.02$ & $0.034\pm0.003$ & 6,10\tabularnewline
\object{131670} & $2930\pm12$ & $0.16\pm0.01$ & $0.040\pm0.002$ & 4,6\tabularnewline
\object{-01$^\circ$3022} & $3253\pm31$ & $0.28\pm0.02$ & $0.016\pm0.001$ & 4\tabularnewline
\object{-14$^\circ$2678} & $3470\pm107$ & $0.22\pm0.04$ & $0.023\pm0.003$ & 4\tabularnewline
\object{59852} & $3464\pm54$ & $0.15\pm0.06$ & $0.0022\pm0.0004$ & 4\tabularnewline
\object{180622} & $4049\pm38$ & $0.06\pm0.10$ & $0.07\pm0.02$ & 0\tabularnewline
\object{216219} & $4098\pm111$ & $0.10\pm0.04$ & $0.013\pm0.001$ &5\tabularnewline
\object{183915} & $4341\pm25$  & $0.47\pm0.06$  & $0.00009\pm0.00004$ & 0\tabularnewline
\object{165141} & $4760\pm120$ & $0.05\pm0.03$ & $0.015\pm0.001$ &  20\tabularnewline
\object{-10$^\circ$4311} & $4888.\pm14.$ & $0.078\pm0.007$ & $0.062\pm0.004$ & 0\tabularnewline
\object{199394} & $5205\pm7$ & $0.14\pm0.01$ & $0.039\pm0.001$ & 0,22\tabularnewline
\object{139195} & $5324.\pm19.$ & $0.35\pm0.02$ & $0.026\pm0.002$ & 12\tabularnewline
\object{40430} & 5570: & 0.25: & - & 0\tabularnewline
\object{22589} & $5761\pm88$ & $0.28\pm0.03$ & $0.0030\pm0.0004$ & 0\tabularnewline
\object{202109} & $6489\pm31$ & $0.22\pm0.03$ & $0.023\pm0.003$ & 13 \tabularnewline
\object{196673} & $7636\pm153$ & $0.66\pm0.03$ & $0.019\pm0.006$ & 0\tabularnewline
\object{18182} & 8056: & 0.3: & - & 0\tabularnewline
\object{53199} & $8300.\pm99.$ & $0.24\pm0.01$ & $0.029\pm0.003$ & 0\tabularnewline
\object{134698} & 9386: & 0.91: & - & 0\tabularnewline
\object{51959} & 9488: & 0.58: & - &0\tabularnewline
\object{104979} & 13940: & 0.2: & - & 0\tabularnewline
\object{98839} & $16419\pm116$ & $0.556\pm0.006$& $0.058\pm0.004$ & 0\tabularnewline
\object{119185} & 19467: & 0.4: & - & 0\tabularnewline
\object{50843} & SB &  & & 0\tabularnewline
\object{95345} & cst? &  & & 0\tabularnewline
\hline
\end{tabular}
\end{table*}

\begin{table*}[ht]
\caption{\label{Tab:S}
Same as Table~\ref{Tab:strongBa} for S (and C) stars.}
\begin{tabular}{lllllllll}
\multicolumn{1}{c}{Name (HD/DM)} & \multicolumn{1}{c}{Period (d)} & \multicolumn{1}{c}{$e$} & 
\multicolumn{1}{c}{$f(M)$ (M$_{\odot}$)} & Ref.\\
\hline\\
\noalign{Extrinsic S stars} 
\medskip\\
\object{121447} & $185.7\pm0.1$ & $0.01\pm0.01$ & $0.025\pm0.001$ & 1\tabularnewline
\object{95875} & $197.2\pm0.3$ & 0.0 & $0.061\pm0.007$ & 19 (Hen4-108) \tabularnewline
\object{-25$^\circ$10393} & $346.6\pm0.03$ & $0.104\pm0.003$ & $0.066\pm0.001$ & 0,19 (Hen 4-147)\\
\object{191589} & $377.3\pm0.1$ & $0.250\pm0.003$ & $0.394\pm0.005$ & 4 \tabularnewline
\object{189581} & $614\pm1$  & $0.39\pm0.02$ & $(1.3\pm0.1)\times 10^{-5}$ & 0\\ 
\object{22649} & $596.2\pm0.2$ & $0.09\pm0.02$ & $0.037\pm0.003$ & 15 \tabularnewline
\object{35155} & $638.2\pm0.3$ & $0.07\pm0.03$ & $0.032\pm0.003$ & 16 (V1261 Ori) \tabularnewline
\object{332077} & $669.1\pm1.0$ & $0.077\pm0.007$ & $1.25\pm0.02$ & 10 \tabularnewline
\object{+24$^\circ$620} & $773.4\pm5.5$ & $0.06\pm0.03$ & $0.042\pm0.005$ & 10 \tabularnewline
\object{+22$^\circ$700} & $849.5\pm8.8$ & $0.08\pm0.06$ & $0.043\pm0.008$ & 10\tabularnewline
\object{+23$^\circ$3093} & $1008.1\pm4.8$ & $0.39\pm0.03$ & $0.045\pm0.005$ & 10\tabularnewline
\object{215336} & $1143.3\pm0.6$ & $0.021\pm0.003$ & $0.039\pm0.003$ & 0\\
\object{9810} & $1147\pm2$ & $0.21\pm0.01$ & $0.019\pm0.001$ & 19 (Hen 4-2)\\
\object{63733} & $1160.7\pm8.9$ & $0.23\pm0.03$ & $0.025\pm0.003$ & 4 \tabularnewline
\object{191226} & $1210.4\pm4.3$ & $0.19\pm0.02$ & $0.013\pm0.001$ & 17\tabularnewline
\object{+28$^\circ$4592} & $1252.9\pm3.5$ & $0.09\pm0.02$ & $0.016\pm0.001$ & 4 \tabularnewline
\object{246818} & $2548.5\pm73.2$ & $0.18\pm0.11$ & $0.004\pm0.002$ & 4\tabularnewline
\object{49368} & $2996\pm67$ & $0.36\pm0.05$ & $0.022\pm0.003$ & 4\tabularnewline
\object{+23$^\circ$3992} & $3096\pm42$ & $0.10\pm0.03$ & $0.034\pm0.004$ & 4 \tabularnewline
\object{343486} & $3166\pm38$ & $0.24\pm0.03$ & $0.039\pm0.005$ & 4\tabularnewline
\object{+21$^\circ$255}(S) & $4137\pm317$ & $0.21\pm0.04$ & $0.032\pm0.004$ & 4\tabularnewline
\object{170970} & $4651\pm10$ & $0.19\pm0.01$ & $0.0213\pm0.0007$ & 0\tabularnewline
\object{+31$^{\circ}$4391} & $6757\pm37$ & $0.14\pm0.03$ & $0.03\pm0.01$ & 0\tabularnewline
\object{+79$^\circ$156} & $11119\pm69$ & $0.44\pm0.01$ & $0.022\pm0.002$ & 0 \tabularnewline
\object{184185} & 17490: & 0.37: & - & 0\tabularnewline
\object{288833} & 17565: & 0.42: & - & 0\tabularnewline
\object{218634} & 90359: & 0.55: & - & 0\tabularnewline
\medskip\\
\noalign{Symbiotic S and C stars} \\
\object{-28$^\circ$3719} & $399.1\pm0.1$ & $<0.002$ & $0.020\pm0.001$ & 0 (Hen 4-18)\tabularnewline 
\object{V420 Hya} & $750.0\pm0.2$ & $0.092\pm0.002$ & $0.0928\pm0.0004$ & 0,19 (Hen 4-121)\tabularnewline
\object{59643} &  $1305\pm4$ & $0.18\pm0.03$ & $0.027\pm0.003$ & 18 (C)\\
\object{ER Del} & $2056\pm2$ & $0.233\pm0.004$ & $0.069\pm0.001$ & 16\\
\object{7351} & $4559\pm9$ & $0.11\pm0.01$ & $0.062\pm0.001$ & 0\tabularnewline
\hline
\medskip\\
\end{tabular}

References to Tables~\ref{Tab:strongBa}--\ref{Tab:S}: (0) Jorissen et al., in preparation; (1) \citet{1995A&A...301..707J}; (2) \citet{2008Obs...128..474G};
(3) \citet{2011AJ....141..144S}; (4)  \citet{1998A&AS..131...25U}; (5) \citet{1998A&AS..131...43U};  (6) \citet{1990ApJ...352..709M}; (7) \citet{2007A&A...473..829M}; (8) \citet{1980MNRAS.193..957G}; (9) \citet{1996Obs...116..298G}; (10) \citet{1998A&A...332..877J};
(11) \citet{2006Obs...126....1G};  (12) \citet{1991Obs...111...29G}; (13) \citet{1992Obs...112..168G}; (14) \citet{2009Obs...129....6G}; (15) \citet{1984Obs...104..224G}; (16) \citet{2014A&A...564A...1B};  (17) \citet{1998A&AS..131...49C}; (18) \citet{2008AN....329...44C}; (19) \citet{2000A&AS..145...51V}; (20) Fekel, priv. comm.; (21) A new orbital solution has been computed from \citet{1990ApJ...352..709M} data, and the non-zero eccentricity has been adopted from that new computation (despite being compatible with a circular orbit); (22) A solution with $P =$ 10480~d, $e = $0.36, and $f(M) = 0.11\pm0.01$~M$_{\odot}$  is also possible, although less probable given the large value of the mass function.

\end{table*}

\begin{table*}[h]
   \centering
      \caption{\label{Tab:cluster_mass}
Open cluster stars and derived masses.}
   %\topcaption{Table captions are better up top} % requires the topcapt package
   \begin{tabular}{lrrcl} % Column formatting, @{} suppresses leading/trailing space
\hline\hline
      Open Cluster  & Star & \multicolumn{1}{c}{$f(m)$} & Mass & Sigma \\
                            &        &      \multicolumn{1}{c}{(M$_\odot$)} &  (M$_\odot$) & (M$_\odot$) \\
      \hline
IC4651  & 6686  &  0.060700  &  2.050  &  0.05 \\
IC4651  & 8665 &  0.290000  &  2.050  &  0.05 \\
IC4651  & 10195 &  0.010380  &  2.050  &  0.05 \\
IC4651  & 14290 &  0.142000  &  2.050  &  0.05 \\
IC4651  & 14641 &  0.004820  &  2.050  &  0.05 \\
IC4725  & 150  &  0.349000  &  5.47  &  0.05 \\
IC4756  & 69  &  0.023700  &  2.630  &  0.05 \\
IC4756  & 80 &  0.203000  &  2.630  &  0.05 \\
IC4756  & 139 &  0.006450  &  2.630  &  0.05 \\
Mel25  & 41 &  0.001010  &  2.480  &  0.05 \\
Mel25  & 71 &  0.170000  &  2.480  &  0.05 \\
Mel71  & 107 &  0.399000  &  3.467  &  0.05 \\
Mel71  & 110 &  0.255400  &  3.467  &  0.05 \\
Mel71  & 118 &  0.231700  &  3.467  &  0.05 \\
Mel71  & 151 &  0.167300  &  3.467  &  0.05 \\
Mel105  & 17 &  0.040000  &  3.800  &  0.05 \\
Mel111  & 91 &  0.260000  &  2.760  &  0.05 \\
NGC0129  & 170 &  0.914000  &  5.90  &  0.05 \\
NGC0129  & 200 &  0.267500  &  5.90  &  0.05 \\
NGC0752  & 75 &  0.090300  &  1.940  &  0.05 \\
NGC0752  & 110 &  0.218600  &  1.940  &  0.05 \\
NGC0752  & 208 &  0.080400  &  1.940  &  0.05 \\
NGC1027  & 27 &  0.079600  &  4.330  &  0.11 \\
NGC1528  & 4 &  0.907000  &  2.978  &  0.05 \\
NGC1778  & 2 &  0.001210  &  3.000  &  0.05 \\
NGC1817  & 44 &  0.065400  &  2.770  &  0.05 \\
NGC1817  & 56 &  0.011700  &  2.770  &  0.05 \\
NGC1817  & 164 &  0.305400  &  2.764  &  0.05 \\
NGC1817  & 244 &  0.000440  &  2.764  &  0.05 \\
NGC2099  & 49 &  0.040100  &  3.300  &  0.05 \\
NGC2099  & 149 &  0.011040  &  3.300  &  0.05 \\
NGC2099  & 485 &  0.188400  &  3.350  &  0.05 \\
NGC2099  & 748 &  0.560000  &  3.084  &  0.05 \\
NGC2099  & 782 &  0.011200  &  3.120  &  0.05 \\
NGC2099  & 966 &  0.011	   & 3.05     & 0.05 \\
NGC2215  & 26 &  0.023000  &  3.585  &  0.05 \\
NGC2287  & 21 &  0.016400  &  3.803  &  0.05 \\
NGC2287  & 97 &  0.000027  &  3.795  &  0.05 \\
NGC2287  & 102 &  0.670000  &  3.541  &  0.05 \\
NGC2287  & 107 &  0.037800  &  3.795  &  0.05 \\
NGC2324  & 1006 &  0.192000  &  2.700  &  0.1 \\
NGC2335  & 4 &  0.015250  &  4.400  &  0.05 \\
NGC2360  & 44 &  0.185000  &  2.600  &  0.13 \\
NGC2360  & 51 &  0.072400  &  2.730  &  0.05 \\
NGC2360  & 52 &  0.039800  &  2.480  &  0.05 \\
NGC2360  & 62 &  0.011900  &  2.480  &  0.05 \\
NGC2360  & 181 &  0.474000  &  2.600  &  0.13 \\
NGC2420  & 173 &  0.007900  &  2.096  &  0.05 \\
NGC2420  & 250 &  0.047300  &  2.096  &  0.05 \\
NGC2423  & 43 &  0.000850  &  2.320  &  0.05 \\
NGC2437  & 29 &  0.018900  &  3.535  &  0.05 \\
NGC2437  & 242 &  0.322300  &  3.588  &  0.05 \\
NGC2447  & 25 &  0.208900  &  2.926  &  0.05 \\
NGC2447  & 42 &  0.041000  &  2.926  &  0.05 \\
     \hline
   \end{tabular}
  \label{tab:mass}
\end{table*}

\addtocounter{table}{-1}

\begin{table*}[htbp]
   \centering
      \caption{Open cluster stars and derived masses (cont.).}
   %\topcaption{Table captions are better up top} % requires the topcapt package
   \begin{tabular}{lrrcl} % Column formatting, @{} suppresses leading/trailing space
\hline\hline
      Open Cluster  & Star &  \multicolumn{1}{c}{$f(m)$}  & Mass & Sigma \\
                           &        &      \multicolumn{1}{c}{(M$_\odot$)} &  (M$_\odot$) & (M$_\odot$) \\
      \hline
NGC2477  & 1025 &  0.006980  &  2.330  &  0.05 \\
NGC2477  & 1044 &  0.010700  &  2.330  &  0.05 \\
NGC2477  & 1272 &  0.000048  &  2.330  &  0.05 \\
NGC2477  & 2064 &  0.081900  &  2.330  &  0.05 \\
NGC2477  & 2204 &  0.545000  &  2.330  &  0.05 \\
NGC2477  & 3003 &  0.209000  &  2.330  &  0.05 \\
NGC2477  & 3170 &  0.013500  &  2.330  &  0.05 \\
NGC2477  & 3176 &  0.012860  &  2.330  &  0.05 \\
NGC2477  & 4067 &  0.000220  &  2.330  &  0.05 \\
NGC2477  & 4137 &  0.095200  &  2.330  &  0.05 \\
NGC2477  & 5073 &  0.057900  &  2.330  &  0.05 \\
NGC2477  & 6020 &  0.015400  &  2.330  &  0.05 \\
NGC2477  & 6062 &  0.011800  &  2.330  &  0.05 \\
NGC2477  & 6251 &  0.000300  &  2.330  &  0.05 \\
NGC2477  & 8017 &  0.064300  &  2.330  &  0.05 \\
NGC2477  & 8018 &  0.004010  &  2.330  &  0.05 \\
NGC2482  & 23 &  0.071700  &  3.000  &  0.10 \\
NGC2533  & 17  &  0.001940  &  2.400  &  0.125 \\
NGC2539  & 114 &  0.012400  &  3.000  &  0.05 \\
NGC2539  & 209 &  0.017200  &  3.000  &  0.05 \\
NGC2539  & 209 &  0.015400  &  3.000  &  0.05 \\
NGC2539  & 223 &  0.351000  &  3.000  &  0.05 \\
NGC2539  & 233 &  0.000013  &  3.000  &  0.05 \\
NGC2539  & 663 &  0.001170  &  3.000  &  0.05 \\
NGC2548  & 1296 &  0.007900  &  3.035  &  0.05 \\
NGC2548  & 1560 &  0.547000  &  3.150  &  0.15 \\
NGC2567  & 104 &  0.280300  &  3.330  &  0.08 \\
NGC2632  & 428 &  0.018980  &  2.410  &  0.08 \\
NGC2682  & 136 &  0.002300  &  1.502  &  0.05 \\
NGC2682  & 143 &  0.002680  &  1.511  &  0.05 \\
NGC2682  & 170 &  0.005900  &  1.555  &  0.05 \\
NGC2682  & 224 &  0.142000  &  1.567  &  0.05 \\
NGC2682  & 244 &  0.009100  &  1.522  &  0.05 \\
NGC2972  & 14 &  0.008120  &  5.450  &  0.11 \\
NGC3532  & 152 &  0.172800  &  3.187  &  0.05 \\
NGC3532  & 160 &  0.010030  &  3.330  &  0.05 \\
NGC3680  & 27 &  0.006100  &  2.000  &  0.10 \\
NGC3960  & 50 &  0.070800  &  2.370  &  0.05 \\
NGC3960  & 91 &  0.030100  &  2.370  &  0.05 \\
NGC3960  & 275 &  0.067600  &  2.45  &  0.13 \\
NGC4349  & 79 &  0.085200  &  3.860  &  0.08 \\
NGC4349  & 203 &  0.016510  &  3.860  &  0.08 \\
NGC5822  & 2 &  0.009160  &  2.490  &  0.11 \\
NGC5822  & 3 &  0.023000  &  2.490  &  0.11 \\
NGC5822  & 4 &  0.024200  &  2.490  &  0.11 \\
NGC5822  & 11 &  0.039400  &  2.490  &  0.11 \\
NGC5822  & 80 &  0.087430  &  2.490  &  0.11 \\
NGC5822  & 151 &  0.011700  &  2.490  &  0.11 \\
NGC5822  & 276 &  0.004600  &  2.490  &  0.11 \\
NGC5822  & 312 &  0.196000  &  2.490  &  0.11 \\
NGC5823  & 1034 &  0.288000  &  2.200  &  0.20 \\
NGC6124  & 29 &  0.006260  &  4.570  &  0.11 \\
NGC6124  & 33 &  0.358000  &  4.570  &  0.11 \\
     \hline
   \end{tabular}
\end{table*}

\addtocounter{table}{-1}

\begin{table*}[htbp]
   \centering
      \caption{Open cluster stars and derived masses (cont.).}
   %\topcaption{Table captions are better up top} % requires the topcapt package
   \begin{tabular}{lrrcl} % Column formatting, @{} suppresses leading/trailing space
\hline\hline
      Open Cluster  & Star &  \multicolumn{1}{c}{$f(m)$}  & Mass & Sigma \\
                           &        &      \multicolumn{1}{c}{(M$_\odot$)} &  (M$_\odot$) & (M$_\odot$) \\
      \hline
NGC6134  & 8 &  0.034100  &  2.250  &  0.11 \\
NGC6134  & 34 &  0.014200  &  2.250  &  0.11 \\
NGC6134  & 204 &  0.005310  &  2.250  &  0.11 \\
NGC6192  & 96 &  0.000550  &  4.53  &  0.05 \\
NGC6475  & 58 &  0.025600  &  3.500  &  0.05 \\
NGC6475  & 134 &  0.017990  &  3.500  &  0.05 \\
NGC6633  & 70 &  0.807000  &  2.820  &  0.05 \\
NGC6694  & 14 &  0.001080  &  5.560  &  0.05 \\
NGC6705  & 926 &  0.750000  &  3.840  &  0.05 \\
NGC6705  & 1223 &  0.430000  &  3.840  &  0.05 \\
NGC6709  & 303 &  0.150800  &  4.430  &  0.11 \\
NGC6940  & 84 &  0.031100  &  2.308  &  0.05 \\
NGC6940  & 92 &  0.005300  &  2.312  &  0.05 \\
NGC6940  & 100 &  0.181200  &  2.306  &  0.05 \\
NGC6940  & 111 &  0.020500  &  2.312  &  0.05 \\
NGC6940  & 130 &  0.027600  &  2.312  &  0.05 \\
NGC6940  & 189 &  0.280000  &  2.308  &  0.05 \\
NGC7209  & 95 &  0.018130  &  2.980  &  0.13 \\
     \hline
   \end{tabular}
\end{table*}

\section{Chemical analysis}
\label{Sect:abundances_appendix}
\subsection{Stellar parameters}
\label{Sect:parameters}
The effective temperature $T_{\mathrm{eff}}$ was derived from the de-reddened $(V-K)_0$ index, 
using the photometric data listed in the WEBDA database,
and the calibration of  \citet{1998A&A...333..231B}, which also provided the bolometric correction.
The surface gravity $\log g$ was derived from its definition $g = G M / R^2$, with $G$ the gravitational constant, $M$ the stellar mass (taken from Table~\ref{Tab:cluster_mass}), and $R$ the stellar radius, derived from the Stefan-Boltzmann relation, using $T_{\rm eff}$ as above, and $L$ from the cluster distance modulus and bolometric magnitude. The cluster metallicity, as listed in Table~\ref{Tab:targets}, is adopted from the WEBDA database.
The microturbulence velocity $v_{\mathrm{turb}}$ was set at 1.5~km~s$^{-1}$ for all stars.

\subsection{Line list}
We compiled the atomic line list from the line database VALD\footnote{\url{http://vald.astro.univie.ac.at/~vald/php/vald.php}} \citep{1999A&AS..138..119K, 2000BaltA...9..590K}. Molecular transitions from ${}^{12}\mathrm{C}{}^{14}\mathrm{N}$ and ${}^{13}\mathrm{C}{}^{14}\mathrm{N}$  (Plez, priv. comm.) were  also included in the computation of synthetic spectra. We chose the lines to be measured following the prescription by \citet{2013A&A...560A..44V}, \citet{2014A&A...567A..30M}, and \citet{2016A&A...586A.151M}. We took into account the hyperfine structure for the following lines: Ba~II at 585.3668~nm, 614.1711~nm, and 649.6900~nm, and La~II at 626.2422~nm, and 639.0477~nm. Table~\ref{Tab:linelists} provides the final list of lines used in this paper.

\begin{table*}
  \begin{center}
    \caption{\label{Tab:linelists} List of measured lines: wavelength $\lambda$, excitation potential $\chi_{\rm exc}$, adopted oscillator strengths $\log gf$, and source:  (a) \citet{2016A&A...586A.151M}, (b) GES line list v5 \citep{2015PhyS...90e4010H}, (c) adjusted to reproduce solar and/or Arcturus abundances, and  (d) VALD. The effective $\log gf$ is given for lines whose hyperfine structure was taken into account (identified with an asterisk after the wavelength). }
\begin{tabular}{lllrl}
  \hline
  \hline
  \noalign{\smallskip}
  Species & \multicolumn{1}{c}{$\lambda$}        & \multicolumn{1}{c}{{$\chi_{\rm exc}$} }          & {$\log gf_{\mathrm{adopted}}$} \\
          &  \multicolumn{1}{c}{(\AA)} &  \multicolumn{1}{c}{(eV)} &                            \\
  \hline
  \noalign{\smallskip}
\ion{Y}{II}     & 4883.682  & 1.084 & 0.000         & c \\  
\ion{Y}{II}     & 5087.416  & 1.084 & -0.330    & c \\    
\ion{Y}{II}     & 5200.406  & 0.992 & -0.730    & c \\    
\ion{Y}{II}     & 5289.815  & 1.033 & -1.900    & c \\    
\ion{Y}{II}     & 5320.782  & 1.084 & -1.950    & d \\    
\ion{Y}{II}     & 5402.774  & 1.839 & -0.640    & c \\    
\ion{Y}{II}     & 5544.611  & 1.738 & -0.980    & c \\    
\ion{Y}{II}     & 5546.009  & 1.748 & -1.120    & c \\    
\ion{Y}{II}     & 5728.887  & 1.839 & -1.300    & c \\    
\ion{Y}{II}     & 6795.414  & 1.738 & -1.620    & c \\    
\ion{Zr}{I}         & 4772.310  & 0.623 & 0.040     & d \\
\ion{Zr}{I}         & 6127.440  & 0.154 & -1.060        & d \\
\ion{Zr}{I}         & 6134.550  & 0.000 & -1.280        & d \\
\ion{Zr}{I}         & 6143.200  & 0.071 & -1.100        & d \\
\ion{Zr}{II}    & 5112.270  & 1.665 & -0.850    & d \\
\ion{La}{II}    & 4558.460  & 0.321 & -0.970    & d \\
\ion{La}{II}    & 4574.860  & 0.173 & -1.080    & d \\
\ion{La}{II}    & 4662.500  & 0.000 & -1.240    & d \\
\ion{La}{II}    & 4748.730  & 0.927 & -0.540    & d \\
\ion{La}{II}    & 4804.039* & 0.235 & -1.490    & b \\
\ion{La}{II}    & 4920.980  & 0.126 & -0.580    & d \\
\ion{La}{II}    & 5114.560  & 0.235 & -1.030    & d \\
\ion{La}{II}    & 5290.820  & 0.000 & -1.650    & d \\
\ion{La}{II}    & 5303.528* & 0.321 & -1.350    & b \\
\ion{La}{II}    & 5797.570  & 0.244 & -1.360    & d \\
\ion{La}{II}    & 5880.630  & 0.235 & -1.830    & d \\
\ion{La}{II}    & 6390.478* & 0.321 & -1.41     & b \\
\ion{La}{II}    & 6774.268  & 0.126 & -1.820    & b \\
\ion{Ce}{II}    & 4515.849  & 1.058 & -0.240    & d \\
\ion{Ce}{II}    & 4523.075  & 0.516 & -0.240    & c \\
\ion{Ce}{II}    & 4562.359  & 0.478 & 0.210         & b \\
\ion{Ce}{II}    & 4628.239  & 1.366 & -0.430    & d \\
\ion{Ce}{II}    & 4773.941  & 0.924 & -0.390    & b \\
\ion{Ce}{II}    & 5274.229  & 1.044 & -0.170    & c \\
\ion{Ce}{II}    & 5472.279  & 1.247 & -0.100    & d \\
\ion{Ce}{II}    & 5975.818  & 1.327 & -0.450    & b \\
\ion{Ce}{II}    & 6043.373  & 1.206 & -0.480    & a \\
\ion{Nd}{II}    & 4645.760  & 0.559 & -0.760    & d \\
\hline\\
\end{tabular}

 \end{center}
\end{table*}

\subsection{Abundance determination}
We use spectral synthesis, as described in \cite{2013A&A...560A..44V}, to derive the chemical abundances of the \emph{s}-elements Y, Zr, La, Ce, and Nd. To this end, a grid of theoretical spectra is computed with \emph{turbospectrum}  \citep{1998A&A...330.1109A,Plez2012}, while stellar-atmosphere models are interpolated from the grid of MARCS (spherical) model atmospheres\footnote{models available at \url{http://marcs.astro.uu.se/}} \citep{2008A&A...486..951G}. 

\begin{figure}
    \includegraphics[width=0.9\columnwidth]{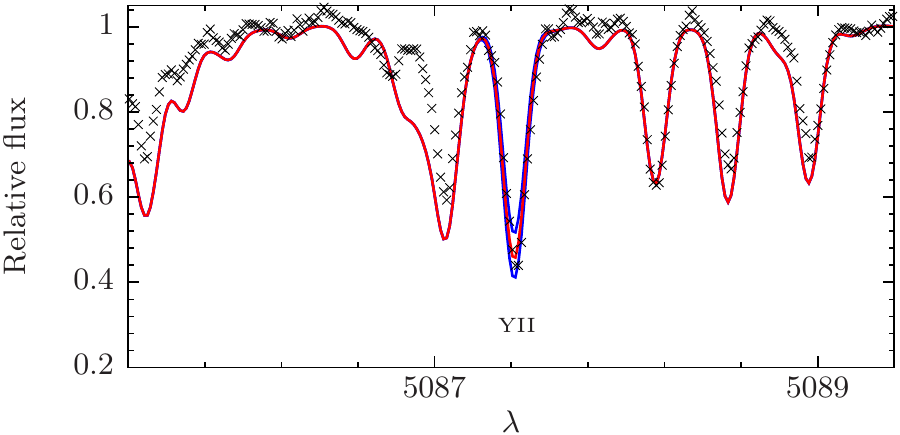}

    \includegraphics[width=0.9\columnwidth]{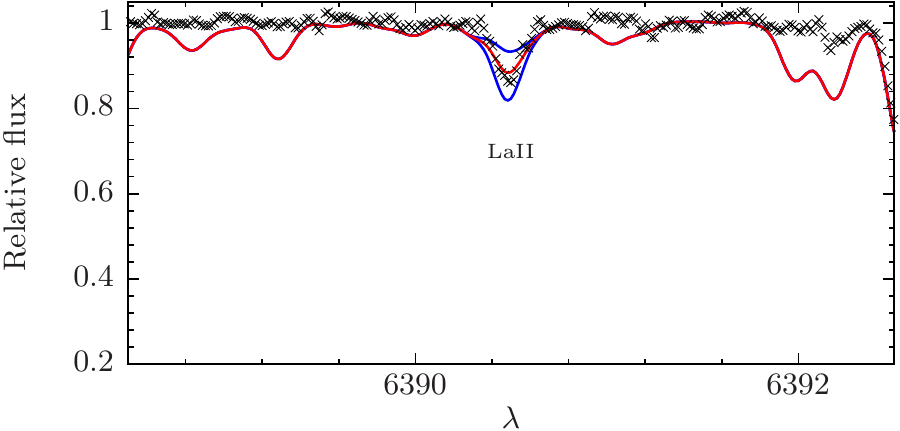}

    \includegraphics[width=0.9\columnwidth]{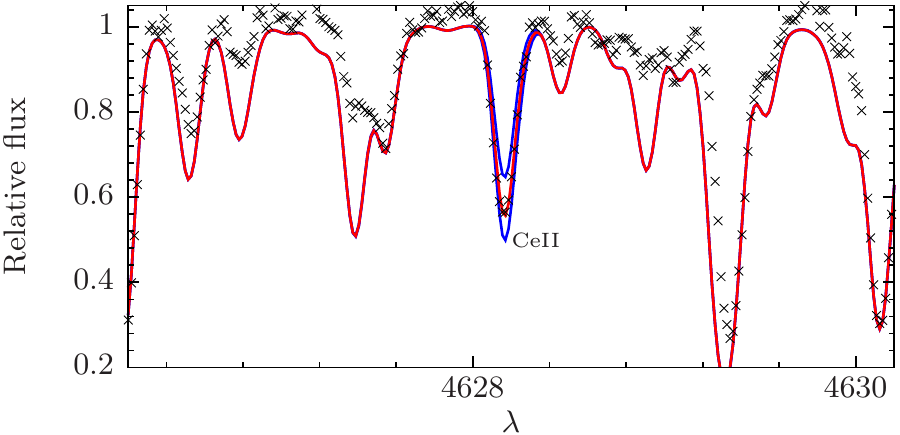}

    \includegraphics[width=0.9\columnwidth]{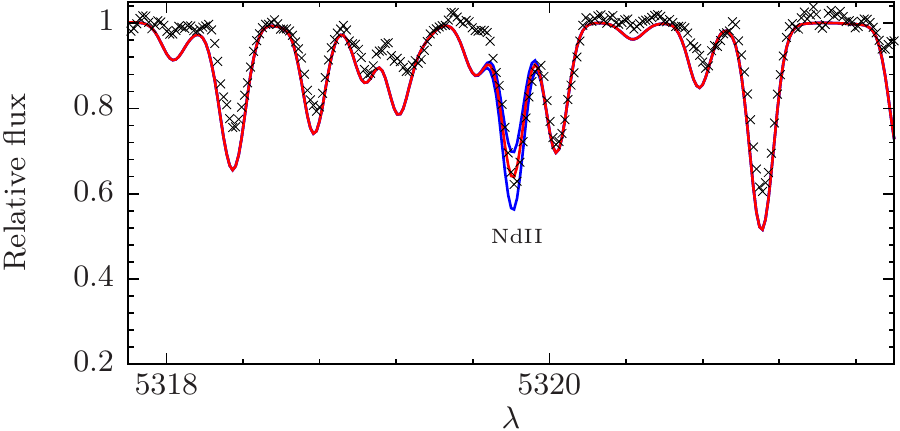}

\caption[]{\label{Fig:example_fit} 
From top to bottom: fit of the  \ion{Y}{II} line at 508.7~nm, 
\ion{La}{II} line at 639.0~nm, \ion{Ce}{II} line at 462.8~nm, and \ion{Nd}{II} line at 531.9~nm 
for IC~4756.69. Black crosses: observed spectrum; red solid line: best fit; blue solid lines: spectra computed around the best fit value [X/Fe]$ \pm 0.3$~dex.
}
\end{figure}

The radiative transfer is computed at local thermodynamical equilibrium (LTE) in spherical geometry. As described in \citet{2013A&A...560A..44V}, the $\chi^{2}$ algorithm selecting the best-fit model weights each wavelength by the total flux of the contaminating species, such that the more contaminated a pixel is, the less it counts in the $\chi^{2}$. The contaminating flux is obtained from a synthetic spectrum where the line under consideration has been suppressed, thus revealing the contribution from all other species.
Figure~\ref{Fig:example_fit} shows the fits to the observed spectrum of the star IC~4756.69. Tables~\ref{Tab:final_abundances} and \ref{Tab:line-by-line_abundances_all_clusters}  give averaged  and line-by-line abundances for the target stars.

\begin{table*}
  \begin{center}
    \caption{\label{Tab:line-by-line_abundances_all_clusters} Line-by-line $\abratio{X}{Fe}$ abundance ratios for Y, Zr, La, Ce, and Nd for open cluster stars falling in the post-mass-transfer region of the ($P-e$) diagram. The cluster metallicities are adopted from the WEDBA database. }
    \begin{tabular}{rrrrrrrrrrrrrrrrrr}
\hline\\
Ion  & \multicolumn{1}{c}{$\lambda$}& 
                    Sun		    &IC4756 139	    &NGC2099 149	&NGC2099 966	&NGC2335 4	    &NGC2420 173	&NGC2539 209    \\
\hline\\

[Fe/H]  &           & 0.0           &-0.06          &+0.08          &+0.08          &-0.03          &-0.26          &+0.13         \\
\medskip\\
%Sr	II	&4215.52	&+0.00	     	& --  	     	& --  	     	& --  	     	& --  	     	& --  	     	& --  	  \\
Y 	II	&4883.68	&-0.03	     	&+0.06	     	&+0.12	     	&+0.17	     	&+0.79	     	&+0.99	     	&-0.20	    \\ 	     	
%Y 	II	&4900.12	&-0.22	     	&-0.34	     	&-0.29	     	&-0.24	     	&+0.67	     	&+0.38	     	&-0.40	     \\	     	
Y 	II	&5087.41	&+0.00	     	&-0.25	     	&-0.09	     	&-0.28	     	&+1.28	     	&+1.65	     	&-0.11	     	\\     	
Y 	II	&5200.41	&+0.00	     	& --  	     	&+0.02	     	& --  	     	&+1.28	     	&+1.15	     	&-0.18	     	  \\   	
Y 	II	&5289.82	&+0.00	     	& --  	     	& --  	     	& --  	     	&+0.63	     	& --  	     	& --  	     	    \\ 	
Y 	II	&5320.78	& --  	     	&+0.05	     	&-0.16	     	&-0.14	     	& --  	     	&+0.62	     	&-0.04	     	 \\    	
Y 	II	&5402.77	&+0.01	     	&-0.04	     	& --  	     	&-0.25	     	&-0.02	     	&+1.04	     	&-0.04	     	  \\   	
Y 	II	&5544.61	&+0.00	     	&+0.07	     	&-0.09	     	&+0.04	     	&-0.06	     	&+0.88	     	&+0.02	     	   \\  	
Y 	II	&5546.01	&+0.01	     	&+0.12	     	&+0.01	     	&-0.08	     	&-0.14	     	&+0.88	     	& --  	     	    \\ 	
Y 	II	&5728.89	&+0.01	     	&+0.09	     	&+0.13	     	&+0.32	     	&+0.36	     	&+1.04	     	&+0.24	     	    \\ 	
Y 	II	&6795.41	&-0.05	     	&+0.07	     	&+0.05	     	&+0.09	     	&+0.04	     	&+0.74	     	&+0.16	     	    \\ 	
{\bf [Y II/Fe]}& &{\bf-0.01}        &{\bf+0.02}     &{\bf+0.00}     &{\bf-0.02}    &{\bf+0.46}      &{\bf+1.00}     &{\bf-0.02}      \\
               & &${\bf\pm0.02}$    &${\bf\pm0.11}$ &${\bf\pm0.10}$ &${\bf\pm0.20}$&${\bf\pm0.53}$  &${\bf\pm0.28}$ &${\bf\pm0.15}$  \\
\medskip\\
Zr	I 	&4772.32	&+0.10	     	& --  	     	& --  	     	&-0.22	     	& --  	     	& --  	     	&-0.35	  \\   		     	
Zr	I 	&6127.48	&-0.04	     	&+0.43	     	&+0.01	     	&-0.16	     	& --  	     	&+0.72	     	&-0.30	  \\   		     	
Zr	I 	&6134.59	&+0.02	     	&+0.26	     	& --  	     	& --  	     	& --  	     	&+0.71	     	& --  	  \\   		     	
Zr	I 	&6143.25	& --  	     	&+0.32	     	&-0.07	     	&-0.21	     	& --  	     	&+0.72	     	& --  	  \\   		     	
{\bf [Zr I/Fe]}& 	&{\bf+0.03}     &{\bf+0.34}     &{\bf-0.03}     &{\bf-0.20}     & --  	   	    &{\bf+0.72}     &{\bf-0.32}      \\
         &          &${\bf\pm0.06}$ &${\bf\pm0.07}$ &${\bf\pm0.04}$ &${\bf\pm0.03}$ &               &${\bf\pm0.00}$ &${\bf\pm0.02}$  \\
\medskip\\
{\bf [Zr II/Fe]}& 5112.27 &{\bf-0.01} & --  	    &{\bf-0.18}      & --  	   	    & --  	    	& --  	   	    &{\bf-0.10}		     	\\
\medskip\\
La	II	&4558.46	&-0.02	     	& --  	     	& --  	     	& --  	     	& --  	     	& --  	     	& --  \\	     	   
La	II	&4574.86	&+0.00	     	&+0.19	     	&+0.31	     	&+0.18	     	& --  	     	&+0.47	     	&-0.20	\\     	     	
La	II	&4662.49	&+0.17	     	&+0.47	     	& --  	     	&+0.26	     	& --  	     	& --  	     	&+0.00	\\     	     	
La	II	&4748.73	&+0.11	     	&+0.20	     	& --  	     	&+0.13	     	&+0.20	     	&+0.71	     	&+0.07	\\     	     	
La	II	&4804.04	& --  	     	&+0.25	     	&+0.13	     	&+0.21	     	& --  	     	&+0.69	     	&+0.18	\\     	     	
La	II	&4920.98	&-0.08	     	&+0.47	     	& --  	     	& --  	     	& --  	     	& --  	     	& --  	\\     	     	
La	II	&5114.56	& --  	     	&+0.50	     	&+0.00	     	&+0.55	     	& --  	     	&+0.86	     	&+0.20\\	     	   
La	II	&5290.82	&+0.02	     	& --  	     	&-0.15	     	&+0.21	     	&+0.22	     	& --  	     	&-0.02\\	     	   
La	II	&5303.53	&-0.03	     	&+0.27	     	&-0.05	     	&+0.03	     	&+0.21	     	&+0.39	     	&+0.08	\\     	     	
La	II	&5797.57	& --  	     	&+0.37	     	&+0.06	     	&-0.07	     	& --  	     	&+0.62	     	& --  \\	     	   
La	II	&5880.63	& --  	     	&+0.38	     	&+0.10	     	&+0.06	     	&+0.21	     	& --  	     	& --  \\	     	     
%La	II	&6320.38	&-0.38	     	&+0.20	     	&-0.12	     	&+0.22	     	&+0.55	     	& --  	     	& --  \\	     	     
La	II	&6390.50	&-0.04	     	&+0.37	     	&+0.02	     	&+0.21	     	&+0.02	     	&+0.58	     	&+0.11\\	     	     
La	II	&6774.27	& --  	     	&+0.35	     	&+0.29	     	&+0.39	     	& --  	     	& --  	     	&+0.17\\	     	     
{\bf [LaII/Fe]}&	&{\bf+0.04}     &{\bf+0.35}      &{\bf+0.08}     &{\bf+0.20}     &{\bf+0.17}     &{\bf+0.62}     &{\bf+0.07}      \\
         &          &${\bf\pm0.07}$ &${\bf\pm0.10}$ &${\bf\pm0.14}$ &${\bf\pm0.16}$ &${\bf\pm0.08}$ &${\bf\pm0.15}$ &${\bf\pm0.12}$  \\
         \medskip\\
Ce	II	&4515.85	&+0.13	     	&+0.00	     	& --  	     	&-0.01	     	& --  	     	& --  	     	& --  	\\     		     	
Ce	II	&4523.08	&+0.00	     	&+0.19	     	&-0.12	     	&-0.18	     	& --  	     	&+0.56	     	& --  	\\     		     	
Ce	II	&4562.36	&-0.11	     	&+0.02	     	&+0.03	     	& --  	     	& --  	     	& --  	     	&-0.29	\\     		     	
Ce	II	&4628.24	& --  	     	& --  	     	& --  	     	& --  	     	& --  	     	& --  	     	& --  	\\     		     	
Ce	II	&4773.94	& --  	     	&+0.24	     	&+0.22	     	& --  	     	&+0.35	     	& --  	     	&+0.27	\\     		     	
Ce	II	&5274.23	&+0.01	     	&+0.31	     	&+0.37	     	&+0.24	     	&+0.33	     	&+0.75	     	&+0.10	\\     		     	
Ce	II	&5472.28	& --  	     	&+0.03	     	&-0.17	     	&-0.29	     	&-0.20	     	& --  	     	&-0.15	\\     		     	
Ce	II	&5975.82	& --  	     	&+0.42	     	&+0.16	     	&+0.17	     	&+0.12	     	& --  	     	&+0.10	\\     		     	
Ce	II	&6043.37	&-0.16	     	&+0.27	     	&+0.05	     	&+0.11	     	& --  	     	&+0.47	     	&+0.09	\\     		     	
{\bf [CeII/Fe]}&	&{\bf-0.03}     &{\bf+0.18}     &{\bf+0.08}     &{\bf+0.01}     &{\bf+0.15}     &{\bf+0.59}     &{\bf+0.02}      \\
               &    &${\bf\pm0.10}$ &${\bf\pm0.14}$ &${\bf\pm0.18}$ &${\bf\pm0.19}$ &${\bf\pm0.22}$ &${\bf\pm0.12}$ &${\bf\pm0.18}$  \\
\medskip\\
{\bf [NdII/Fe]}&4645.76 & --  	   	&{\bf+0.36}	    &{\bf+0.06}	    &{\bf+0.08}	    &{\bf+0.63}	    &{\bf+1.51}     & -- 	\\           \\
\hline\\
\end{tabular}

  \end{center}
\end{table*}

\addtocounter{table}{-1}
\begin{table*}
  \begin{center}
    \caption{Continued.}

\begin{tabular}{rrrrrrrrrrrrrrrrrr}
\hline\\
Ion  & \multicolumn{1}{c}{$\lambda$} &NGC2682 143     &NGC2682 170   &NGC2682 244    &NGC6940 111 \\
\hline\\

[Fe/H]  &         &+0.00                     &+0.00          &+0.00          &+0.01      \\
\medskip\\
%Sr	II	&4215.52& --  	& --  	     	& --  	     	& --  \\
Y 	II	&4883.68&+0.01	& --  	     	&-0.31	     	&-0.05\\
%Y 	II	&4900.12& --  	& --  	     	&-0.58	     	&-0.15\\
Y 	II	&5087.41& --  	&-0.59	     	&-0.34	     	&-0.18\\
Y 	II	&5200.41&-0.31	& --  	     	&-0.44	     	&-0.23\\
Y 	II	&5289.82&+0.09	&-0.03	     	&-0.03	     	& --  \\
Y 	II	&5320.78& --  	&-0.23	     	&-0.07	     	&-0.05\\
Y 	II	&5402.77&-0.13	& --  	     	&-0.19	     	&+0.07\\
Y 	II	&5544.61& --  	&-0.02	     	&+0.06	     	&+0.17\\
Y 	II	&5546.01& --  	& --  	     	&+0.11	     	&+0.10\\
Y 	II	&5728.89&+0.44	&+0.17	     	&-0.03	     	&+0.28\\
Y 	II	&6795.41&+0.26	&+0.05	     	&+0.06	     	&+0.17\\
{\bf [Y II/Fe]}& &{\bf+0.06}    &{\bf-0.11}     &{\bf-0.12}     &{\bf+0.03}\\
               & &${\bf\pm0.25}$&${\bf\pm0.25}$ &${\bf\pm0.18}$ &${\bf\pm0.16}$
\medskip\\
Zr	I 	&4772.32	&+0.12&-0.36	     	&+0.10	     	& --  \\
Zr	I 	&6127.48	&+0.09&-0.09	     	&+0.32	     	&+0.49\\
Zr	I 	&6134.59	&+0.30&-0.22	     	&+0.17	     	&+0.37\\
Zr	I 	&6143.25	& --  &-0.36	     	&+0.09	     	&+0.32\\
{\bf [Zr I/Fe]}& &{\bf+0.17}    	&{\bf-0.26}     &{\bf+0.17}     &{\bf+0.39}\\
         &       &${\bf\pm0.09}$ &${\bf\pm0.11}$ &${\bf\pm0.09}$ &${\bf\pm0.07}$
\medskip\\
{\bf [Zr II/Fe]}& 5112.27 & --  	&{\bf-0.23}		&{\bf-0.28}     &{\bf-0.10} \\
\medskip\\
La	II	&4558.46	& --  	& --  	     	& --  	     	&-0.18\\
La	II	&4574.86	& --  	&-0.24	     	&-0.19	     	&+0.11\\
La	II	&4662.49	& --  	&-0.16	     	&-0.11	     	&+0.11\\
La	II	&4748.73	&+0.03	&+0.02	     	&-0.17	     	&+0.15\\
La	II	&4804.04	& --  	&-0.11	     	&-0.04	     	&+0.17\\
La	II	&4920.98	& --  	& --  	     	& --  	     	&+0.18\\
La	II	&5114.56	& --  	&-0.02	     	&-0.03	     	&+0.23\\
La	II	&5290.82	&+0.42	&-0.07	     	&-0.16	     	&+0.18\\
La	II	&5303.53	&+0.20	&-0.13	     	&-0.16	     	& --  \\
La	II	&5797.57	& --  	& --  	     	&+0.02	     	&+0.20\\
La	II	&5880.63	& --  	& --  	     	& --  	     	&+0.28\\
%La	II	&6320.38	& --  	&-0.11	     	&-0.05	     	&+0.17\\
La	II	&6390.50	&+0.28	&-0.04	     	&+0.07	     	&+0.30\\
La	II	&6774.27	& --  	&-0.13	     	& --  	     	&+0.39\\
{\bf [LaII/Fe]}&   &{\bf+0.23}    &{\bf-0.10}     &{\bf-0.08}     &{\bf+0.18}\\
         &        &${\bf\pm0.14}$ &${\bf\pm0.07}$ &${\bf\pm0.08}$ &${\bf\pm0.13}$
\medskip\\
Ce	II	&4515.85	& --  & --  	     	& --  	     	& --  \\
Ce	II	&4523.08	&-0.11& --  	     	&-0.35	     	&-0.23\\
Ce	II	&4562.36	&-0.26& --  	     	&-0.44	     	&-0.09\\
Ce	II	&4628.24	& --  & --  	     	& --  	     	& --  \\
Ce	II	&4773.94	&+0.37&-0.13	     	&+0.24	     	&+0.31\\
Ce	II	&5274.23	&+0.04&-0.39	     	&-0.01	     	&+0.29\\
Ce	II	&5472.28	& --  &-0.46	     	& --  	     	&-0.13\\
Ce	II	&5975.82	& --  & --  	     	&+0.16	     	&+0.32\\
Ce	II	&6043.37	&+0.18&-0.17	     	&+0.05	     	&+0.24\\
{\bf [CeII/Fe]}&   &{\bf+0.04}    &{\bf-0.29}     &{\bf-0.06}     &{\bf+0.10}\\
               &  &${\bf\pm0.22}$&${\bf\pm0.14}$ &${\bf\pm0.25}$ &${\bf\pm0.22}$
\medskip\\
{\bf [NdII/Fe]}&4645.76&{\bf+0.22}	 & --  	        &{\bf+0.06} 	&{\bf+0.21}\\
\hline\\
\end{tabular}

  \end{center}
\end{table*}

\addtocounter{table}{-1}
\begin{table*}
  \begin{center}
    \caption{Continued.}

\begin{tabular}{rrrrrrrrrrrrrrrrrr}
\hline\\
Ion  & \multicolumn{1}{c}{$\lambda$} &IC 4756 69       &IC 4756 80     &NGC 2477 1044   &NGC 2682 224    &NGC 4349 203 & NGC 752 208 \\
\hline\\

[Fe/H]  &        &-0.06          &-0.06          &+0.01          &+0.00          & -0.07   & -0.08      \\
\medskip\\
Y 	II&	4883.68	 &-0.05	     	 &-0.26	     	 &+0.52	     	&-0.13	     	&+0.26	     	&-0.13\\
Y 	II&	4900.12	 & -	     	 & -	     	 &+0.17	     	&-0.51	     	&-0.02	     	&-0.37\\
Y 	II&	5087.41	 &-0.03	     	 & -	     	 &+0.36	     	& -	     	    &+0.13	     	&-0.29\\
Y 	II& 5200.41	 & -	     	 & -	     	 &+0.47	     	& -	     	    &+0.16	     	& -   \\
Y 	II&	5289.82	 &+0.09	     	 &+0.08	     	 &+0.57	     	&-0.04	     	&+0.07	     	&+0.05\\
Y 	II&	5402.77	 &+0.09	     	 &+0.01	     	 &+0.55	     	&+0.01	     	&+0.02	     	&+0.01\\
Y 	II&	5544.61	 &+0.08	     	 &-0.01	     	 &+0.54	     	&+0.07	     	&+0.05	     	&+0.08\\
Y 	II&	5728.89	 &+0.26	     	 &+0.18	     	 &+0.72	     	&+0.23	     	&+0.19	     	&+0.22\\
Y 	II&	6795.41	 &+0.18	     	 &+0.18	     	 &+0.58	     	&+0.01	     	&+0.09	     	&+0.10\\
%\medskip\\
{\bf [Y II/Fe]} &&${\bf +0.09}$&${\bf +0.03}$&${\bf +0.54}$&${\bf +0.03}$&${\bf +0.12}$&${\bf +0.01}$\\
                &&${\bf \pm0.10}$&${\bf \pm0.15}$&${\bf \pm0.10}$&${\bf \pm0.11}$&${\bf \pm0.07}$&${\bf \pm0.15}$\\
\medskip\\
Zr	I&	4739.48	 & -	     	& -	     	&-0.06	     	&-0.22	     	&-0.39	     	&-0.21\\
Zr	I&	4772.32	 &-0.38	     	&-0.30	    &+0.04	     	&-0.04	     	&-0.21	     	&-0.08\\
Zr	I&	5385.15	 & -	     	& -	     	&+0.07	     	&-0.01	     	& -	     		&-0.07\\
Zr	I&	5735.69	 &+0.79	     	&+0.80	    &+1.05	     	&+1.06	     	&+0.97	     	&+0.91\\
Zr	I&	6127.48	 &-0.18	     	&-0.19	    &+0.23	     	& -	     		&-0.08	     	& -\\
Zr	I&	6134.59	 &-0.32	     	&-0.39	    &+0.21	     	&+0.14	     	&-0.14	     	&+0.09\\
Zr	I&	6140.54	 & -	     	& -	     	&+0.48	     	& -	     		& -	     		&+0.27\\
Zr	I&	6143.25	 & -	     	& -	     	&+0.22	     	&+0.10	     	& -	     		&+0.01\\
Zr	I&	6445.75	 & -	     	& -	     	&+0.33	     	&+0.20	     	& -	     		& -\\
Zr	I&	6990.87	 & -	     	& -	     	& -	     		&+0.35	     	& -	     		&+0.83\\
%\medskip\\
{\bf [Zr I/Fe]}  &&${\bf -0.02}$ &${\bf -0.02}$&${\bf +0.29}$&${\bf +0.20}$	&${\bf +0.03}$	&${\bf +0.22}$	\\
	             &&${\bf \pm0.47}$&${\bf \pm0.48}$&${\bf \pm0.31}$&${\bf \pm0.36}$&${\bf \pm0.48}$&${\bf \pm0.40}$
\medskip\\
Zr	II&	5112.27	&+0.12	     	&+0.01	     	&+0.45	     	&-0.16	     	&+0.04	     	&-0.03\\
Zr	II&	5350.09	&+0.35	     	&+0.21	     	&+0.75	     	&+0.16	     	&+0.37	     	&+0.22\\
Zr	II&	5350.35	&+0.35	     	&+0.21	     	&+0.75	     	&+0.17	     	&+0.37	     	&+0.23\\
%\medskip\\
{\bf [Zr II/Fe]} &&${\bf +0.27}$ &${\bf +0.14}$ &${\bf +0.65}$ &${\bf +0.06}$ 	&${\bf +0.26}$ 	&${\bf +0.14}$ \\
                 &&${\bf \pm0.11}$&${\bf \pm0.09}$&${\bf \pm0.14}$&${\bf \pm0.15}$   &${\bf \pm0.16}$&${\bf \pm0.12}$\\
\medskip\\
La	II&	4558.46	&+0.07	     	& -	     		&+0.17	     	& -	     		&-0.02	     	&+0.00\\
La	II&	4574.86	&+0.10	     	&-0.08	     	&+0.47	     	&+0.16	     	&+0.23	     	&+0.27\\
La	II&	4662.49	&+0.15	     	&-0.02	     	&+0.26	     	& -	     		&+0.09	     	&+0.09\\
La	II&	4748.73	&+0.10	     	&+0.01	     	&+0.20	     	&-0.11	     	&+0.01	     	&-0.00\\
La	II&	4804.04	&+0.16	     	&+0.15	     	&+0.36	     	&+0.03	     	&+0.05	     	&+0.19\\
La	II&	4920.98	& -	     		& -	     		& -	     		&+0.10	     	&+0.02	     	&+0.16\\
La	II&	5114.56	&+0.27	     	&+0.20	     	&+0.62	     	&+0.19	     	&+0.42	     	&+0.33\\
La	II&	5290.82	&-0.05	     	&-0.16	     	&+0.21	     	&+0.01	     	&-0.04	     	&+0.07\\
La	II&	5303.53	&+0.05	     	&+0.05	     	&+0.34	     	&+0.00	     	&+0.05	     	&+0.09\\
La	II&	5880.63	& -	     		& -	     		&+0.45	     	&+0.01	     	&-0.05	     	&+0.02\\
La	II&	5936.20	& -	     		& -	     		&+0.24	     	&+0.13	     	& -	     		&+0.22\\
La	II&	6172.70	&+0.20	     	&+0.19	     	&+0.54	     	&+0.40	     	&+0.21	     	&+0.26\\
%La	II&	6320.38	&+0.12	     	&+0.15	     	&+0.35	     	&+0.05	     	&+0.21	     	&+0.19\\
La	II&	6390.50	&+0.10	     	&+0.12	     	&+0.39	     	&+0.11	     	&+0.09	     	&+0.16\\
La	II&	6774.27	&+0.15	     	&+0.18	     	&+0.46	     	&+0.07	     	&+0.23	     	&+0.20\\
%\medskip\\
{\bf [La II/Fe]}&&${\bf +0.12}$  &${\bf +0.07}$ &${\bf +0.36}$	&${\bf +0.09}$	&${\bf +0.11}$	&${\bf +0.15}$\\
                &&${\bf \pm0.08}$& ${\bf \pm0.12}$&${\bf \pm0.13}$ &${\bf \pm0.12}$ &${\bf \pm0.13}$ &${\bf \pm0.10}$\\
\hline
\end{tabular}
  \end{center}
\end{table*}

\addtocounter{table}{-1}
\begin{table*}
  \begin{center}
    \caption{Continued.}
    \begin{tabular}{rrrrrrrrrrrrrrrrrr}
\hline\\
Ion  & \multicolumn{1}{c}{$\lambda$} &IC 4756 69       &IC 4756 80     &NGC 2477 1044   &NGC 2682 224    &NGC 4349 203 & NGC 752 208 \\
\hline\\
\medskip\\
Ce	II&	4515.85	&-0.23	     	&-0.36	     	&+0.07	     	&-0.22	     	&-0.21	     	&-0.18\\
Ce	II&	4523.08	&-0.04	     	&-0.11	     	&+0.15	     	&-0.13	     	&-0.08	     	&-0.04\\
Ce	II&	4628.16	&-0.14	     	&-0.27	     	&-0.02	     	&-0.33	     	&-0.10	     	&-0.21\\
Ce	II&	4773.94	&+0.21	     	&+0.20	     	&+0.41	     	&+0.27	     	&+0.11	     	&+0.29\\
Ce	II&	5187.46	&-0.14	     	&-0.14	     	&-0.03	     	&-0.24	     	&-0.13	     	&-0.13\\
Ce	II&	5274.23	&+0.13	     	&+0.15	     	&+0.28	     	&-0.06	     	&+0.19	     	&+0.11\\
Ce	II&	5330.57	&-0.06	     	&-0.03	     	& -	     		&-0.14	     	& -	     		&-0.06\\
Ce	II&	5472.28	&-0.09	     	&-0.10	     	&+0.06	     	&-0.19	     	&-0.24	     	&-0.14\\
Ce	II&	6043.37	&+0.09	     	&+0.13	     	&+0.34	     	&+0.05	     	&+0.03	     	&+0.11\\
Ce	II&	6051.82	& -	     		& -	     		&+0.15	     	&-0.22	     	& -	     		&-0.15\\
%\medskip\\
{\bf [Ce II/Fe]}&&${\bf -0.03}$	&${\bf -0.06}$	&${\bf +0.16}$	&${\bf -0.12}$	&${\bf -0.05}$	&${\bf -0.04}$\\
                &&${\bf \pm0.14}$ &${\bf \pm0.18}$ &${\bf \pm0.15}$ &${\bf \pm0.16}$ &${\bf \pm0.14}$ &${\bf \pm0.15}$\\
\medskip\\
{\bf [Nd II/Fe]}&	4645.76	&{\bf +0.08}	&{\bf +0.10}	&{\bf +0.26}	&{\bf +0.14}	&{\bf +0.05}	&{\bf +0.08}\\
\hline
\end{tabular}

  \end{center}
\end{table*}

\subsection{Error budget}
\label{Sect:error}
Table~\ref{Tab:final_abundances} gives the standard deviation around the mean abundance ratios, for each ion with at least three independently measured lines. Since we are dealing with small samples, an estimate of the standard error on the mean can be obtained with $k_{\alpha, N - 1} \times s / \sqrt{N}$, where $k_{\alpha, N-1}$ is the percentile of the Student's $t$ distribution with $N - 1$ degrees of freedom, such that the probability $P(-k_{\alpha, N-1} < T < k_{\alpha, N-1})$ is $1 - \alpha$. For a two-sided test and $1 - \alpha = 0.70$ (\emph{i.e.} approximately a $1\sigma$ confidence interval), $k_{0.3, 2} = 1.386$, $k_{0.3, 3} = 1.250$, $k_{0.3, 4} = 1.190$, $k_{0.3, 7} = 1.119$, $k_{0.3, 8} = 1.108$. This allows us to estimate a conservative random error for our abundance ratios of  0.1~dex.

Table~\ref{Tab:systematic_errors} facilitates the evaluation of the systematic errors on the final abundance ratios caused by the uncertainty on the stellar parameters. As expected, neutral species are more sensitive to the temperature than ionised species, while the tendency is opposite for gravity. Except for the notable case of Zr~I, the impact of the model-atmosphere uncertainties is always below 0.1~dex, in absolute value, which we adopt as a conservative estimate of the systematic error on the abundances.

\begin{table*}
  \begin{center}
    \caption{\label{Tab:systematic_errors} Abundance uncertainties (in dex) for the star IC~4756.69, for uncertainties of $\Delta T_{\mathrm{eff}} = 150$~K, $\Delta \log g = 0.2$, $\Delta\abratio{M}{H} = 0.1$~dex, and $\Delta v_{\mathrm{t}} = 0.2$~km~s$^{-1}$. The value $\Delta_{\pm}\zeta$ is defined as $\abratio{X}{Fe}(\zeta \pm \Delta \zeta) - \abratio{X}{Fe}(\zeta_{\text{nominal}})$, where X is the element under study and $\zeta$ is one of the four atmospheric parameters.}
    \begin{tabular}{lllllllllllllllllllllll}
%\begin{tabular}{lS[table-format=5.2]S[table-format=5.2]S[table-format=5.2]S[table-format=5.2]S[table-format=5.2]S[table-format=5.2]S[table-format=5.2]S[table-format=5.2]}
  \hline
  \hline
  \noalign{\smallskip}
  Ion         & {$\Delta_{+}(T)$} & {$\Delta_{-}(T)$} & {$\Delta_{+}(\log g)$} & {$\Delta_{-}(\log g)$} & {$\Delta_{+}(\abratio{M}{H})$} & {$\Delta_{-}(\abratio{M}{H})$} & {$\Delta_{+}(v_{\text{t}})$} & {$\Delta_{-}(v_{\text{t}})$} \\
  \hline
  \noalign{\smallskip}
%{$\abratio{Sr~I}{Fe}$}  & +0.20 & -0.24 & +0.02 & -0.04 & -0.14 & +0.14 & -0.14 & +0.16 \\
{$\abratio{Y ~II}{Fe}$} & +0.02 & -0.02 & +0.08 & -0.09 & -0.07 & +0.08 & -0.09 & +0.10 \\
{$\abratio{Zr~I}{Fe}$}  & +0.22 & -0.26 & +0.00 & +0.00 & -0.12 & +0.10 & +0.00 & +0.00 \\
{$\abratio{Zr~II}{Fe}$} & +0.00 & +0.01 & +0.10 & -0.10 & -0.09 & +0.09 & -0.02 & +0.03 \\
%{$\abratio{Ba~II}{Fe}$} & +0.09 & -0.07 & +0.03 & -0.03 & -0.06 & +0.07 & -0.17 & +0.13 \\
{$\abratio{La~II}{Fe}$} & +0.04 & -0.05 & +0.09 & -0.10 & -0.09 & +0.08 & -0.01 & +0.01 \\
{$\abratio{Ce~II}{Fe}$} & +0.04 & -0.09 & +0.09 & -0.12 & -0.08 & +0.06 & -0.05 & +0.09 \\
{$\abratio{Nd~II}{Fe}$} & +0.06 & -0.04 & +0.10 & -0.08 & -0.06 & +0.09 & -0.06 & +0.08 \\
%{$\abratio{Eu~II}{Fe}$} & +0.00 & -0.02 & +0.08 & -0.10 & -0.08 & +0.08 & -0.02 & +0.00 \\
  \noalign{\smallskip}
  \hline                  
\end{tabular}

  \end{center}
\end{table*}

\end{document}